\documentclass[preprint,12pt,3p]{elsarticle}

\usepackage{graphicx}
\usepackage{amssymb}
\usepackage{caption}
\usepackage{subcaption}
\usepackage{url}
\usepackage{amsmath}
\usepackage{float}

\journal{Special issue Biomathematics: Mathematical Methods in the Applied Sciences}

\begin{document}

\begin{frontmatter}

\title{\textit{IB2d} Reloaded: a more powerful Python and MATLAB implementation of the immersed boundary method}

\author[label0,label1,label3]{Nicholas A. Battista\corref{cor1}}
\cortext[cor1]{I am corresponding author}
\address[label0]{Department of Mathematics and Statistics, The College of New Jersey, 2000 Pennington Rd., Ewing, NJ 08628}
\ead{nickabattista@gmail.com}
\ead[url]{nickabattista.wixsite.com/home/}

\address[label1]{Department of Mathematics, CB 3250, University of North Carolina, Chapel Hill, NC, 27599}

\author[label1,label2]{W. Christopher Strickland}
\address[label2]{Department of Mathematics, University of Tennessee at Knoxville, 1403 Circle Drive, Knoxville TN, 37996}
\ead{cstric12@utk.edu}

\author[label1]{Aaron Barrett}
\ead{abarret@live.unc.edu}

\author[label1,label3]{Laura A. Miller}
\address[label3]{Department of Biology, CB 3280, University of North Carolina, Chapel Hill, NC, 27599}
\ead{lam9@unc.edu}

%
%

\begin{abstract}
The immersed boundary method (IB) is an elegant way to fully couple the motion of a fluid and deformations of an immersed elastic structure. In that vein, the \textit{IB2d} software allows for expedited explorations of fluid-structure interaction for beginners and veterans to the field of computational fluid dynamics (CFD). While most open source CFD codes are written in low level programming environments, \textit{IB2d} was specifically written in high-level programming environments to make its accessibility extend beyond scientists with vast programming experience. Although introduced previously in \cite{BattistaIB2d:2016}, many improvements and additions have been made to the software to allow for even more robust models of material properties for the elastic structures, including a data analysis package for both the fluid and immersed structure data, an improved time-stepping scheme for higher accuracy solutions, and functionality for modeling slight fluid density variations as given by the Boussinesq approximation. 

\end{abstract}

%
%
\begin{keyword}
Immersed boundary method \sep fluid-structure interaction \sep mathematical biology \sep biomechanics
\end{keyword}

\end{frontmatter}

%
%

%
%



%
%

\section{Introduction}
\label{introduction}



Fluid-structure interaction models (FSI) are creeping into all disciplines in science \cite{Bathe:2008}. Its applications can range from guiding the engineering design of airplanes, boats, and cars for transportation \cite{Bak:2013,Rao:2015,Vanderhoydonck:2016} to understanding the locomotion of aquatic organisms \cite{Fauci:1988,Hershlag:2011,Hoover:2015} and animal flight \cite{Miller:2004,Ruck:2010,SJones:2015}, or personalized medicine \cite{Wolters:2005,Torii:2008,Takizawa:2011} and surgical planning and practice \cite{Yang:2007,Carson:2017} to understanding the role of blood flow in heart development \cite{Santhanakrishnan:2009,Miller:2011,Battista:2016a,Battista:2016b}. 

Fully coupled FSI does not prescribe the motion of the structure that is immersed in a fluid. The deformations of the structure are due to the movement of the fluid, and the movement of the fluid is induced by the forces exerted by the deformations of the structure. For example, if one models ventricular contraction, they can do it two ways. They could elect to prescribe the motion of the ventricle itself, perhaps basing the movement on kinematic data, and observe the reaction of the fluid to the wall movement. Or they may model the ventricular contraction based on an electrophysiology model that feeds into a muscular force generation model which drives contraction. In this case, the ventricle walls are free to deform under the motion of the fluid. This is fully coupled FSI, where the motion of the ventricular walls are not prescribed, and hence they are not acting as a rigid surface, only moving in a dictated fashion, but are free to deform in reaction to the fluid moving. 

The immersed boundary ($IB$) method provides an intuitive framework to study fully coupled FSI problems. It was first developed by Charles Peskin in 1972 to study blood flow around cardiac valves and in turn, the deformations of valve leaflets due to underlying blood flow \cite{Peskin:1972}. Moreover, it also allows functionality to prescribe motion of immersed structures \cite{BattistaIB2d:2016}. It has been successfully applied to study the FSI between an elastic structure immersed within an incompressible fluid for a variety of biological and engineering applications within the intermediate Reynolds number ($Re$) regime, e.g. $Re\sim \mathcal{O}(0.01,1000)$. $Re$ is given as 
\begin{equation}
\label{Re}Re = \frac{\rho L V}{\mu},
\end{equation}
where $\mu$ and $\rho$ are the dynamic viscosity and density of the fluid, respectively, and $L$ and $V$ are a characteristic length and velocity scale of the problem. It is used to capture the correct biological or engineering scale in fluid dynamic models. 

The elegance behind $IB$ lies in its ability to solve fully coupled fluid-structure interaction problems which involve complicated time-dependent geometries while using a regular fixed Cartesian discretization of the fluid domain. The immersed structure is composed of elastic fibers, which govern its material properties, and is discretized on a Lagrangian mesh, e.g. it is not constrained to a fixed Cartesian grid. The movement of the fluid and elastic fibers are coupled; the immersed structure moves at the local fluid velocity while also applying a singular force onto the fluid arising from deformations of the structure itself.

There are many ways to model the fiber elasticity properties of the immersed structure \cite{Battista:2015, BattistaIB2d:2016} giving rise to a wide range of formidable applications. Some immersed boundary examples that illustrate this variety include cardiovascular dynamics \cite{Peskin:1996,GriffithThesis:2005}, aquatic locomotion \cite{Zhang:2014,Hoover:2017}, insect flight \cite{Miller:2009,SJones:2015}, parachute dynamics \cite{Kim:2006}, muscle-fluid-structure interactions \cite{Tytell:2010,Battista:2015,Hamlet:2015}, plant biomechanics \cite{Zhu:2011,Miller:2012}, soap filaments \cite{Zhu:2002,Zhu:2003}, and cellular and other microscale interactions \cite{Atzberger:2007,Leiderman:2008,Fogelson:2008,Strychalski:2013}. Furthermore, the $IB$ framework invites one to add other constitutive models such as electrophysiology, cellular signaling, or chemical reaction equations into the FSI framework \cite{Kramer:2008,Fogelson:2008,Tytell:2010,Lee:2010,Strychalski:2012,Du:2014,Baird:2015,Waldrop:2015,BattistaIB2d:2016,BattistaBioMath:2017}.

In this paper, we are releasing a more finely-tuned version of \textit{IB2d}, which is $IB$ software with full implementations in both MATLAB \cite{MATLAB:2015a} and Python 3.5 \cite{Python:Python} that is capable of modeling a wide range of fluid-structure interaction applications in engineering or biomechanics. It extends the functionality of the software described in \cite{Battista:2015,BattistaIB2d:2016} by implementing more robust models for modeling material properties of the elastic structures, a data analysis toolbox for analyzing both fluid and immersed structure data, an improved time-stepping scheme for higher accurate FSI solutions, functionality for modeling slight fluid density variations as given by the Boussinesq approximation, and other improvements targeting code transparency, utility, and speed. The updated software package also contains $50$+ examples, some of which come from previous $IB$ literature, which illustrate the breadth of the software.


%
%
%
%
%
%
%
%

\section{IBM Framework}
\label{IBM_Framework}

\textit{IB2d} models two dimensional fluid motion coupled to the motion of an immersed structure moving within it. The fluid motion is governed by the Navier-Stokes equations in Eulerian form, written as

\begin{equation}
\label{Navier_Stokes} \rho \left( \frac{\partial {\bf{u}}({\bf{x}},t) }{\partial t} + {\bf{u}}({\bf{x}},t)\cdot \nabla {\bf{u}}({\bf{x}},t) \right) = -\nabla p({\bf{x}},t) + \mu \Delta {\bf{u}}({\bf{x}},t) + {\bf{f}}({\bf{x}},t) \\
\end{equation}
\begin{equation}
\label{Incompressibility} \nabla\cdot {\bf{u}}({\bf{x}},t) = 0,\\
\end{equation}
where ${\bf{u}}({\bf{x}},t) = (u({\bf{x}},t),v({\bf{x}},t))$ is the fluid velocity, $p({\bf{x}},t)$ is the pressure, and ${\bf{f}}({\bf{x}},t)$ is the force per unit volume (or area in $2D$) applied to the fluid by elastic forces arising arising from deformations of the immersed structure. The independent variables are the position, ${\bf{x}}= (x,y)$, and time, $t$. Eq.(\ref{Navier_Stokes}) is the statement of conservation of momentum for a fluid while Eq.(\ref{Incompressibility}) is the incompressibility condition enforcing mass conservation of the fluid. \textit{IB2d} currently assumes a periodic domain. A future release will include functionality for a projection method fluid solver to enforce Dirichlet and Neumann boundary conditions \cite{Chorin:1967,Brown:2001}.

The interaction equations essentially model all communication between the immersed deformable structure and the fluid and are given by the following integral equations with delta function kernels, 
\begin{equation}
\label{IBM_Force} {\bf{f}}({\bf{x}},t) = \int {\bf{F}}(r,t)\delta({\bf{x}}-{\bf{X}}(r,t)) dr
\end{equation}
\begin{equation}
\label{IBM_Velocity} {\bf{U}}({\bf{X}}(r,t),t) = \frac{\partial {\bf{X}}(r,t)}{\partial t} = \int {\bf{u}}({\bf{x}},t) \delta( {\bf{x}} - {\bf{X}}(r,t) ) d{\bf{x}},
\end{equation}
where ${\bf{X}}(r,t)$ gives the Cartesian coordinates at time $t$ of the material point labeled by Lagrangian parameter $r$, ${\bf{F}}(r,t)$ is the force per unit area imposed by elastic deformations in the immersed structure onto the fluid, as a function of the Lagrangian position, $r$, and time, $t$. The force density, ${\bf{F}}(r,t)$, is a functional of the current immersed boundary's configuration. Moreover, the force density is modeled as
\begin{equation}
\label{IBM_Force_Density} {\bf{F}}(r,t) = \bf{\mathbb{F}}{(\bf{X}}(r,t),t),
\end{equation}
where $\bf{\mathbb{F}}({\bf{X}},t)$ is a combination of all the constitutive fiber model relations, modeling specific material properties of the immersed structure for a particular application. Previous fiber models implemented in \textit{IB2d} are described in \cite{BattistaIB2d:2016}, and new fiber model implementations, e.g. damped springs, non-invariant beams, poroelastic media, coagulation, and user-defined force models are described in Section \ref{FiberModels}. 

Eq.(\ref{IBM_Force}) applies a force arising from deformations of the immersed structure's preferred configuration to the fluid through a delta-kernel integral transformation. Essentially the fluid grid points nearest to the Lagrangian point, where the deformation force is given, feel the largest force and such deformation forces taper off at fluid grid points further away. Eq.(\ref{IBM_Velocity}) sets the velocity of the boundary equal to the local fluid velocity to satisfy the no-slip condition on the immersed structure.

The following regularized delta functions, $\delta_h$, are used upon discretizing Eqs.(\ref{IBM_Force}) and (\ref{IBM_Velocity}),  
\begin{equation}
\label{IBM:delta} \delta_h(\mathbf{x}) = \frac{1}{h^2} \phi\left( \frac{x}{h} \right) \phi\left( \frac{y}{h} \right),
\end{equation}
where $h$ is the fluid grid width and
\begin{equation}
\label{IBM:delta2} \phi(r) = \left\{\begin{array}{c} \frac{1}{4} \left(1 + \cos\left(\frac{\pi r}{2} \right) \right) \ \ \ \ \ \ |r|\leq 2 \\ 0  \ \ \ \ \ \ \ \ \ \  \ \ \ \ \ \ \ \ \ \ \ \ \ \ \ \ \ \ \ \  \mbox{otherwise} \end{array}\right.,
\end{equation}
where $r$ is the distance from the Lagrangian node. This particular regularized delta function has compact support. There exist other discrete delta functions with compact support which have been incorporated into $IB$ frameworks \cite{Peskin:2002}, and these may be easily incorporated into the software if desired. 
More details on regularized delta functions may be found in \cite{Peskin:2002,Liu:2012}. 

In a traditional immersed boundary framework, the coupled equations (\ref{Navier_Stokes}-\ref{IBM_Velocity}) are solved using the algorithm described in Peskin's $IB$ review paper \cite{Peskin:2002} with periodic boundary conditions imposed on both the fluid and immersed boundary. The standard numerical algorithm for immersed boundary \cite{Peskin:2002} is as follows:
\begin{itemize}
\item[\emph{Step 1:}] Compute the force density ${\bf{F}}^{n}(r,t)$ on the immersed boundary from the current boundary deformations, ${\bf{X}}^{n}$, where n indicates the $n^{th}$ time-step. 
\item[\emph{Step 2:}] Use Eq.(\ref{IBM_Force}) to spread these deformation forces from the Lagrangian nodes to the fluid lattice points nearby.
\item[\emph{Step 3:}] Solve the Navier-Stokes equations, Eqs.(\ref{Navier_Stokes}) and (\ref{Incompressibility}), on the Eulerian domain, e.g. update ${\bf{u}}^{n+1}$ and $p^{n+1}$ from ${\bf{u}}^{n}$ and ${\bf{f}}^{n}$. Note: since we are enforcing periodic boundary conditions on the computational domain, the Fast Fourier Transform (FFT) \cite{Cooley:1965,Press:1992} is used to solve for these updated quantities at an accelerated rate.
\item[\emph{Step 4:}] Update the fiber model positions, ${\bf{X}}^{n+1}$, using the local fluid velocities, ${\bf{U}}^{n+1}$, using ${\bf{u}}^{n+1}$ and Eq.(\ref{IBM_Velocity}), e.g. move the immersed structure at the local fluid velocities thereby enforcing \emph{no slip} boundary conditions.
\end{itemize}

The above algorithm had been previously used in \textit{IB2d} and details regarding the discretization for this implementation are found in \cite{BattistaIB2d:2016}. However, in this release we follow the algorithm presented in \cite{Lai:2000}, which gives a formally second-order $IB$ algorithm. Details on this $IB$ algorithm's implementation are found in Section \ref{fluid_additions}.

%
%
%
%
%
%

\section{IB2d New Functionality}

Since the original \textit{IB2d} release papers \cite{Battista:2015,BattistaIB2d:2016}, new functionality has continually been added to the software. In the following sections we we will give an overview of the recent additions to the software for available fiber models and fluid solvers. Later in Section \ref{IB2d_Examples} we will illustrate such functionality through a variety of examples.

%
%
%
%
%
%

\subsection{Fiber Models}
\label{FiberModels}

Since the initial release, \textit{IB2d} has added fiber model functionality for the following:

\begin{enumerate}
\item Damped Springs (ex. tethered ball in channel)
\item Non-invariant Beams (ex. anguilliform swimmer)
\item Poroelastic Media (ex. sea grass in oscillatory flow)
\item Coagulation Model (ex. stirring cells)
\item User-defined Force Model (ex. the rubber-band)
\end{enumerate}


\textbf{Damped Springs} $ $\\

\begin{figure}[H]
	\centering
	\includegraphics[width=0.55\textwidth]{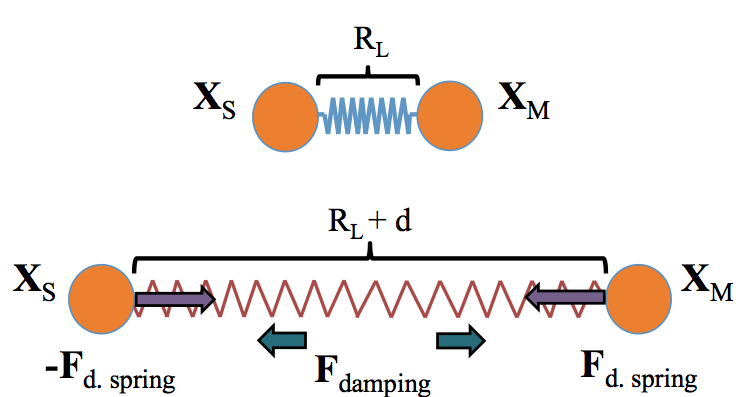}
        	\caption{Illustrated the idea behind the damped springs model. Resistance to stretching or bending is opposed by a friction-like term that is proportional to the velocity of deformation.}
        	\label{fig:D_Springs}
   \end{figure}

Resistance to stretching between successive Lagrangian points can be achieved by modeling the connections with Hookean (or Non-Hookean) springs of resting length $R_L$ and spring stiffness $k_S$. If the virtual spring displacement is below or beyond $R_L$, the model will drive the system back towards a lower energy state, as discussed in \cite{BattistaIB2d:2016}. Moreover, one may also choose to use damped springs, which assumes a frictional damping force that is proportional to the velocity of the oscillation. We note that these take a similar form to the linear spring case, but with an additional term modeling the damping, i.e.,

\begin{equation}
\label{spring:dampingForce} F_{d.spring} = k_S \left( 1 - \frac{R_L}{||{\bf{X}}_{SL}-{\bf{X}}_{M}||} \right) \cdot \left( \begin{array}{c} x_{SL} - x_M \\ y_{SL} - y_M \end{array} \right) + b_S \frac{d}{dt}||{\bf{X}}_{SL}-{\bf{X}}_{M}||,
\end{equation}

where $b_S$ is the damping coefficient. This idea is illustrated in Figure \ref{fig:D_Springs}. \\ $ $\\


\textbf{Non-invariant Beams} $ $\\

\begin{figure}[H]
	\centering
	\includegraphics[width=0.85\textwidth]{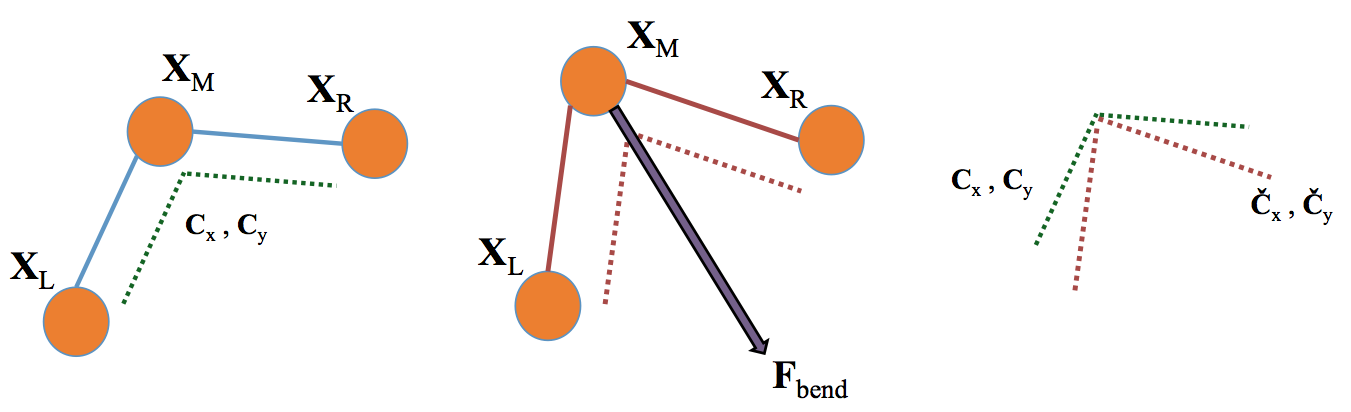}
        	\caption{Motivating the idea of non-invariant beams. The beam has a preferred $x$- and $y$-curvature, given by $C_x and C_y$, respectively. When the configuration is deformed so curvatures are now $\hat{C}_x$ and $\hat{C}_y$, respectively, a restoring force drives the configuration towards its preferred equilibrium position.}
        	\label{fig:NonInv_Beams}
   \end{figure}

Another way to model the resistance bending between three successive Lagrangian points (rather than using torsional springs as in \cite{BattistaIB2d:2016}) is by using a non-invariant beam connecting the three successive nodes. The model assumes a prescribed `curvature' in both $x$ and $y$ components between the three Lagrangian points with corresponding bending stiffness $k_{NIB}$. The corresponding bending deformation forces are modeled as
\begin{equation}
\label{nbeam:force} \mathbf{F}_{beam} = k_{NIB} \frac{\partial^4}{\partial s^4} \left( \textbf{X}(s,t) - \textbf{X}_{b}(s) \right), 
\end{equation}
where $\textbf{X}(s,t)$ is the current Lagrangian configuration at time $t$, e.g. the mapping of the Lagrangian points $s$ to the underlying Cartesian grid, and $\textbf{X}(s)$ is the preferred configuration of the fiber model. More details about this fiber model can be found in \cite{Peskin:2002,Griffith:2009,Bhalla:2013b} and the discretization is discussed below in Section \ref{section:beam}. This fiber model is illustrated in Figure \ref{fig:NonInv_Beams}.

This model is denoted as non-invariant beams since these beams are non-invariant under rotations, as opposed to the torsional spring fiber model. Similarly to the torsional spring model, non-invariant beam deformation forces can only occur on immersed boundary points on the interior or the fiber structure, not the endpoints. 

An example using non-invariant beams is found in Section \ref{IB2d:beam_swimmer}.\\ $ $\\


\textbf{Poroelastic Media} $ $\\

In contrast to porous media being modeled using the traditional form of Darcy's Law such as in previous IB implementations \cite{Kim:2006,Stockie:2009,BattistaIB2d:2016}, one can define a poroelastic structure based on Brinkman-like terms in the momentum equation \ref{Navier_Stokes}. Recall that the traditionally considered Darcy's Law is a phenomenologically derived constitutive equation, which models the fluid velocity through a porous boundary as proportional to the pressure gradient on the two sides of that boundary \cite{Whitaker:1986}. This can be described mathematically as 
\begin{equation}
\label{porosity:darcy} U_p = -\frac{\kappa_p [p]}{\mu \sigma},
\end{equation}
where $U_p \hat{n}$ is the porous slip velocity, $\kappa_p$ is the membrane permeability, $\sigma$ is the structure's thickness, $[p]$ is the pressure gradient across the boundary, and $\hat{n}$ is the unit normal vector to the permeable structure. However, one may also use the Brinkman equations to model the fluid flow through a permeable medium.  The Brinkman equations are a combination of the momentum equation in the Navier-Stokes equations (\ref{Navier_Stokes}) and Darcy's Law (\ref{porosity:darcy}). They account for the dissipation of kinetic energy by viscous shearing and can model the transition between slow flow in porous media, which is governed by DarcyÔøΩs law, and faster flow, as described by (\ref{Navier_Stokes}). The Brinkmann equations take the form
\begin{equation}
\label{porosity:brinkman}  \rho \left( \frac{\partial {\bf{u}}({\bf{x}},t) }{\partial t} + {\bf{u}}({\bf{x}},t)\cdot \nabla {\bf{u}}({\bf{x}},t) \right) = -\nabla p({\bf{x}},t) + \mu \Delta {\bf{u}}({\bf{x}},t) + \mu\alpha(\textbf{x},\textbf{y}) \textbf{u}  + {\bf{f}}({\bf{x}},t),
\end{equation}
where the additional term $\mu\alpha(\textbf{x},\textbf{y}) \textbf{u}$ is the Brinkmann term. $\alpha(\textbf{x},\textbf{y})$ is the inverse of the hydraulic permeability in low $Re$. Note that if a region is not porous, then $\alpha(\textbf{x},\textbf{y})=0$, and if it is porous then $\alpha(\textbf{x},\textbf{y})>0.$ However, rather than add this term into (\ref{Navier_Stokes}) as part of the fluid solve, we will assume that deformations of an immersed elastic structure, composed of springs (whether linear, non-linear, or damped), are balanced by the Brinkmann term, e.g.
\begin{equation}
\label{porosity:balance} \textbf{f}_{brink} = -\textbf{f}_{elastic}.
\end{equation}
\begin{figure}[H]
    \begin{subfigure}{0.5\textwidth}
        \centering
        \includegraphics[width=0.975\textwidth]{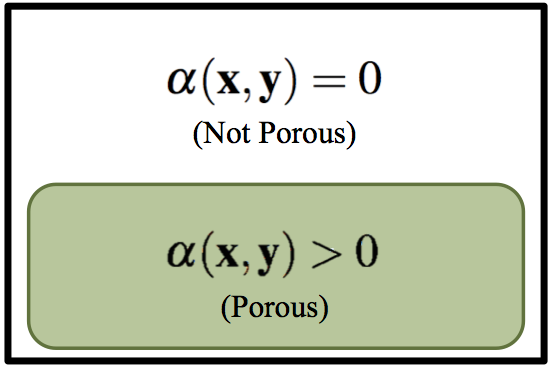}
        \caption{}
        \label{IB_Poro1}
    \end{subfigure}
    \begin{subfigure}{0.5\textwidth}
        \centering
        \includegraphics[width=0.975\textwidth]{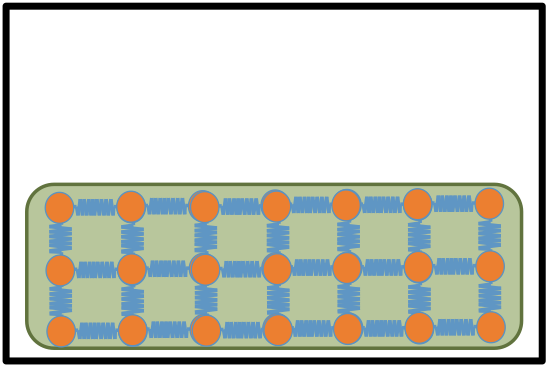}
        \caption{}
        \label{IB_Poro2}
    \end{subfigure}\\ \\
    \caption{Illustrating the idea behind poroelastic media in \textit{IB2d}. (a) Depiction of a region of poroelasticity within a fluid domain. (b) An example of how to construct a poroelastic region with springs. Note you could also attach springs diagonally across, as well.}
    \label{FiberModel:Poroelastic}
\end{figure}

This idea is shown in Figure \ref{FiberModel:Poroelastic}. These elastic deformation forces can then be used to find the slip velocity of the boundary 
\begin{equation}
\label{porosity:slip} \textbf{U}_{b}(\textbf{X},t) = \textbf{u}(\textbf{x},t) + \frac{ \textbf{f}_{elastic} }{ \alpha(\textbf{x},\textbf{y}) \mu}.
\end{equation}
$ $\\


\textbf{Coagulation (Aggregaton) Model} $ $\\

\begin{figure}[H]
	\centering
	\includegraphics[width=0.95\textwidth]{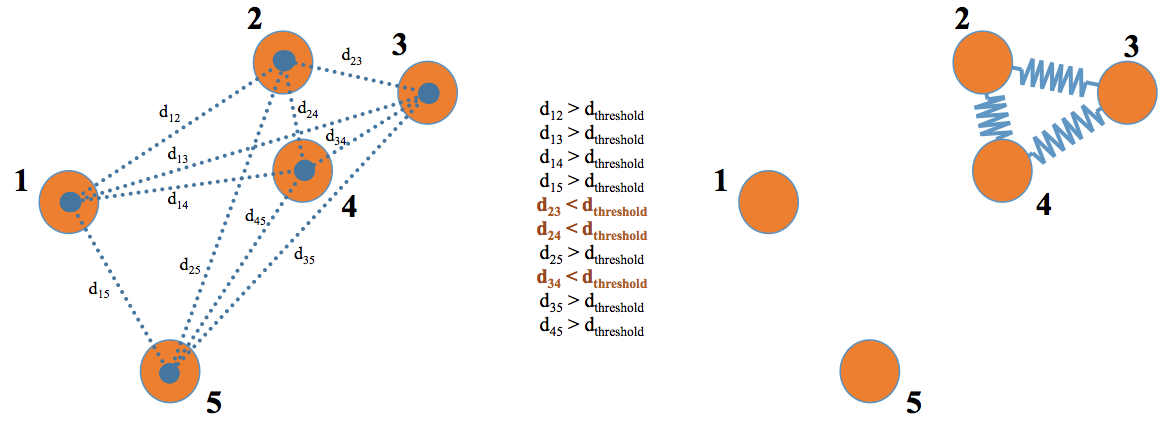}
        	\caption{This diagram illustrates that bonds will be formed between cells, if the distance from cell-center to cell-center is less than some set threshold.}
        	\label{fig:Coag1}
   \end{figure}

One can model dynamically occurring bonds between moving Lagrangian points to model the aggregation of cells or particles. Coagulation and aggregation had been previously introduced into an IB framework in \cite{Fogelson:1984}. Many improvements in modeling blood clotting, coagulation, and thrombus formation in conjunction with flow have been demonstrated in \cite{Fogelson:1992,Fauci:1993,Fogelson:1999,Fogelson:2008,Leiderman:2011,Leiderman:2014}. Our primitive model creates bonds, by initiating a spring connection, between the closest Lagrangian points to its neighboring cell if the cells are within a threshold distance away from one another. This bond (spring) fractures if there is a large enough force to break the bond, set by a fracture threshold.

\begin{figure}[H]
	\centering
	\includegraphics[width=0.95\textwidth]{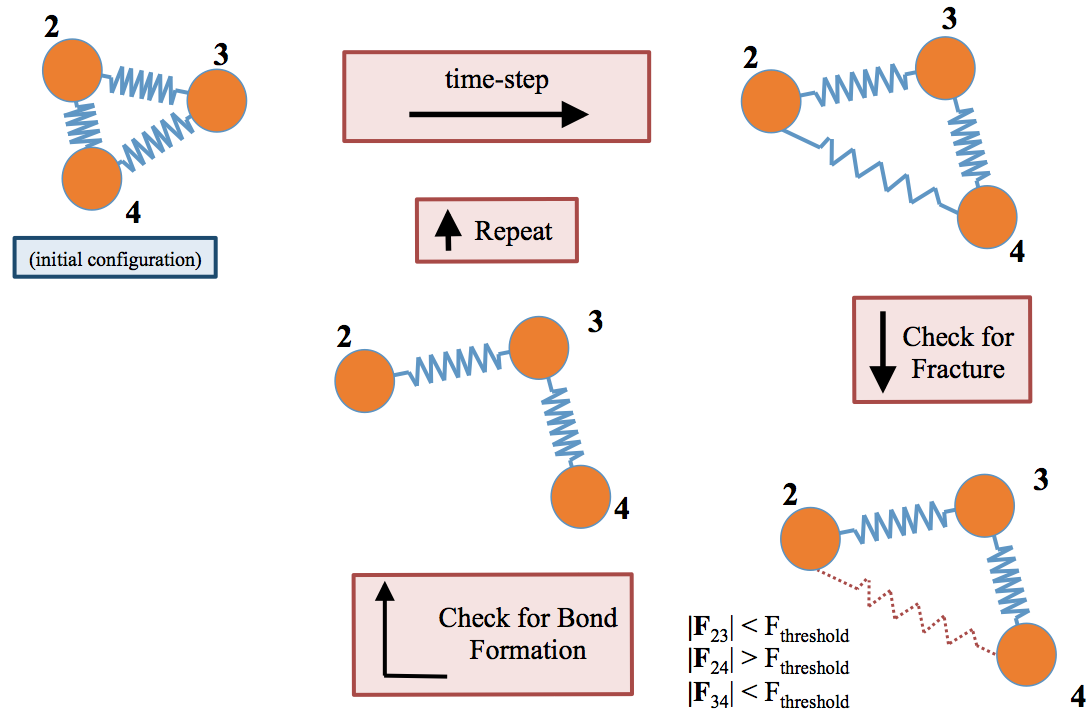}
        	\caption{Schematic of the subset of the IB algorithm that models coagulation. After bonds are formed, the simulation marches forward one time-step, checks for fracture if a bonded force is greater than a threshold value, erases bonds, if necessary, checks for bond formation, and then repeats this process, similarly.}
        	\label{fig:Coag_Timestep}
   \end{figure}

Figures \ref{fig:Coag1} and \ref{fig:Coag_Timestep} illustrate bond formation and bond fracturing for the coagulation and aggregation model. Figure \ref{fig:Coag1} shows that bonds will form when cells are close enough to one enough. This is modeled by a spring (linear, non-linear, damped), where a spring is set between the two Lagrangian Points that are closest together along the neighboring cells. When the bond is formed, the simulation will then proceed forward. After a time-step, it will check for bonds that fracture. If the magnitude of one of the springs is greater than a threshold value, that bond will be erased modeling bond fracture. Next, the algorithm checks for new bond formation and if there are, creates such bonds after which the process repeats. This is shown in Figure \ref{fig:Coag_Timestep}.  \\ $ $\\


\textbf{User-defined Force Model} $ $\\

\textit{IB2d} also allows the option for the user to define their own force model. The user has control in how they define their model as well as what Lagrangian points are involved as many items are passed into the model automatically, such as the current and previous position of the Lagrangian points, current time, time-step, etc, and also includes the functionality for the user to read in appropriately chosen data for parameters, etc. An example of how to use this functionality is shown in Section \ref{section:user_force}.\\


\subsubsection{Using the User-defined Force Model}
\label{section:user_force}
$ $\\

The user-defined force functionality works very similarly to the other fiber models, e.g. there is an associated input file, that data gets read into \textit{IBM\textunderscore Driver} file, and then finally into a function that computes the deformation forces at each time step. However, the difference is that the user defines the style of the input file, and controls the deformation force model script themselves in the particular Example folder.\\

The user-defined force input file ends with a \textit{`struct'.user\textunderscore force}, where \textit{`struct'} is the string name designating that particular example, i.e., the name specified in the \textit{input2d} file. Inside the \textit{.user\textunderscore force} file, the user has the ability to put whatever necessary parameters, Lagrangian IDs, etc., are required for their force model. This is completely analogous to the formats of other input file types. \\

The input file data then gets passed to a script \textit{give\textunderscore Me\textunderscore General\textunderscore User\textunderscore Defined\textunderscore Force\textunderscore Densities}, which needs to be found in the Example folder. That script receives the input parameters listed in Table \ref{User_Defined_Inputs} from the \textit{IBM\textunderscore Driver} file.\\

\begin{table}[H]
\centering
\begin{tabular}{ |r|l| }
  \hline
  \multicolumn{2}{|c|}{Input Parameters for \textit{give\textunderscore Me\textunderscore General\textunderscore User\textunderscore Defined\textunderscore Force\textunderscore Densities}} \\
  \hline
  $ds$ &  Lagragian spacing (defined by $ds = \frac{1}{2}dx$) \\
  $Nb$ & $\#$ of Lagrangian Pts. \\
  $xLag$ & current x-Lagrangian coordinate positions \\
  $yLag$ & current y-Lagrangian coordinate positions \\
  \textit{xLag\textunderscore P} & previous x-Lagrangian coordinate positions \\
  \textit{yLag\textunderscore P} & previous y-Lagrangian coordinate positions \\
  $dt$ & time-step value \\
  \textit{current\textunderscore time} & current time in simulation \\
  \textit{general\textunderscore force} & matrix containing all data from the \textit{.user\textunderscore force} input file \\
  \hline
\end{tabular}
\caption{Descriptions of all the parameters passed into the user-defined deformation force script.}
\label{User_Defined_Inputs}
\end{table}

With these parameters and the data read in from the \textit{.user\textunderscore force} file, the user can define their own Lagrangian deformation force law. 

There is a complete example of this process in the built-in example, \textit{Example\textunderscore User\textunderscore Defined\textunderscore Fiber\textunderscore Model}. This example models an oscillating rubberband in which all successive Lagrangian points are connected by a user-defined deformation force law. For example purposes, the user-defined force here is equivalent to a linear spring. The input files are created by running the \textit{Rubberband} script which generates the \textit{rubberband.vertex} and \textit{rubberband.user\textunderscore force} input files. Note \textit{rubberband.user\textunderscore force} file is equivalent to the input file format for a linear spring. Upon running the simulation, during each time-step the data is passed to the \textit{give\textunderscore Me\textunderscore General\textunderscore User\textunderscore Defined\textunderscore Force\textunderscore Densities} script in which the user defines the actual deformation force model. Here, that script computes a linear spring force between adjacent nodes. This entire process can be summed up as follows:

\begin{enumerate}
\item Generate input files (.vertex, .user\textunderscore force, .etc)
\item Define user-defined force model in the \textit{give\textunderscore Me\textunderscore General\textunderscore User\textunderscore Defined\textunderscore Force\textunderscore Densities} script
\item In the \textit{input2d} file, check the flag for the user-defined force model (and other fiber models, if necessary)
\item Run the simulation
\end{enumerate}


\subsubsection{Discretizing the Fourth Derivatives of the Non-Invariant Beam Model}
\label{section:beam}
$ $\\

Recall that the non-invariant beam deformation forces were given by 
\begin{equation}
\label{app:nbeam:force} \mathbf{F}_{beam} = k_{NIB} \frac{\partial^4}{\partial s^4} \left( \textbf{X}(s,t) - \textbf{X}_{b}(s) \right), 
\end{equation}

First we define
\begin{align}
\nonumber\textbf{X}(s,t) &= (X_q, Y_q), \\
\label{app:beam:lag}\textbf{X}(s+1,t) &= (X_r, Y_r), \\
\nonumber \textbf{X}(s-1,t) &= (X_p, Y_q). 
\end{align}

Recall that by Newton's second law that a force is given by an acceleration, hence we only have to discretize (\ref{app:nbeam:force}) as a second derivative. We find that the forces are computed as
\begin{align}
\nonumber {F}_{beam}(s-1,1) &= - k_{NIB} \Bigg( \begin{array}{l}  X_r - 2X_q + X_p - C_x \\   Y_r - 2Y_q + Y_p - C_y \\ \end{array} \Bigg), \\
\label{app:beam:expForce} {F}_{beam}(s,1) &= 2 k_{NIB}  \Bigg( \begin{array}{l} X_r - 2X_q + X_p - C_x \\   Y_r - 2Y_q + Y_p - C_y \\ \end{array} \Bigg), \\
\nonumber {F}_{beam}(s+1,1) &= -k_{NIB} \Bigg( \begin{array}{l}  X_r - 2X_q + X_p - C_x \\   Y_r - 2Y_q + Y_p - C_y \\ \end{array} \Bigg), 
\end{align}
where $C_x$ and $C_y$ are the preferred curvatures, given by 
\begin{align}
\textbf{C} = \Bigg( \begin{array}{c} C_x \\ C_y \end{array} \Bigg) =  \Bigg( \begin{array}{c} X_{r_{B}} - 2X_{q_{B}} + X_{p_{B}} \\ Y_{r_{B}} - 2Y_{q_{B}} + Y_{p_{B}}  \end{array} \Bigg),
\end{align}
where the $B$ in the subscript denotes the base configuration, or preferred configuration.

%
%
%
%

\subsection{Additions to the fluid solver}
\label{fluid_additions}

Since the initial release of \textit{IB2d}, the fluid solver has been given a few upgrades. First, \textit{IB2d} no longer requires square grids, instead supporting all rectangular grids. Second, a new time-stepping scheme has been implemented that gives rise to a formally second-order accurate IB method \cite{Lai:2000}. Third, functionality for the Boussinesq Approximation \cite{Boussinesq:1897,Tritton:1977} has been added, in conjunction with the advection-diffusion solver. Finally, performance improvements have been added to the Python version of $IB2d$ including an option to leverage the Fastest Fourier Transform in the West (FFTW) \cite{FFTW05} through the pyFFTW library, resulting in an approximate 1.2x speedup over the FFT version. \\


\textbf{Functionality for Rectangular Grids} $ $\\

Previously, \textit{IB2d} only was capable of handling square domains but functionality for rectangular grids has now been incorporated into the framework. For certain applications, this has enormous benefits as it reduces the computational costs, can allow for higher resolution, and reduces the simulation time if a rectangular grid suffices. \\

\begin{figure}[H]
    \centering
    \includegraphics[width=0.9\textwidth]{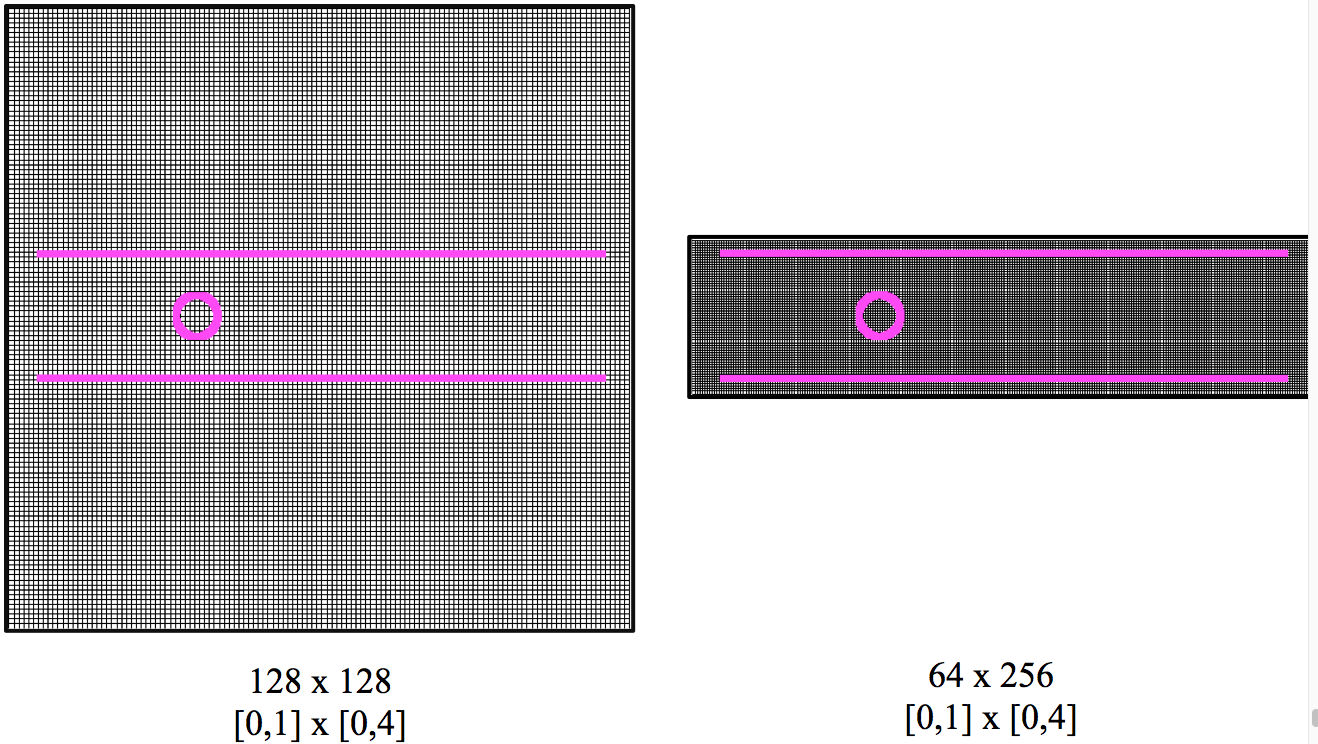}
    \caption{Computational grid illustrating that for approximately the same computational cost of solving a problem on a $128\times128$ grid on a $[0,4]\times[0,4]$ domain, one can solve it on a $64\times256$ grid on a $[0,1]\times[0,4]$ domain. The immersed structure, \textit{shown in pink}, models flow past a cylinder in a channel, which naturally lends itself to rectangular grids.}
    \label{fig:SquareToRectangular}
\end{figure}

For example, if you are modeling a problem in a rectangular channel, this functionality proves particularly beneficial. Consider a domain that is of size $[0,4]m \times [0,4]m$, while a channel that is $3.5m$ long, but only $0.75m$ in diameter (wide). If we wanted $\delta x=\delta y=\frac{4}{128}$ resolution, a square grid would require a $128\times 128$ resolution grid on a $[0,4]m \times [0,4]m$ computational domain. This would require $\sim \mathcal{O}(128^2)=\mathcal{O}(16384)$ operations for the fluid solve each time-step. However, because of the geometry of the structure (a long, narrow channel), and since all the dynamics will happen inside the channel, we can now solve the problem on a computational grid that is $[0,1]m \times [0,1]m$, with $64\times 256$ resolution. This would only require $\sim \mathcal{O}(32\times128)=\mathcal{O}(4096)$ operators for the fluid solve each time-step. In theory this should speed up the $\mathcal{O}(N^2)$ operations by a factor of $4$. Note that these discretizations have the same resolution. Furthermore, for approximately the same computational cost of the $128\times128$ simulation on a $[0,4]m \times [0,4]m$, we could solve the problem on a $64\times256$ grid on a $[0,1]m \times [0,4]m$ domain for twice the computational resolution! This is illustrated in Figure \ref{fig:SquareToRectangular}.

\begin{figure}[H]
    \centering
    \includegraphics[width=0.999\textwidth]{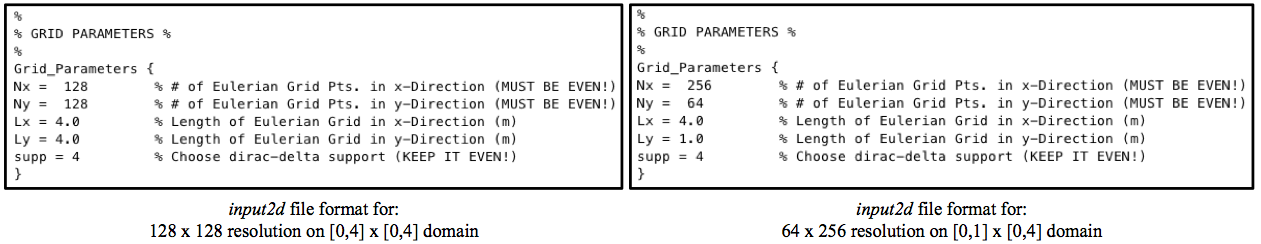}
    \caption{Figure showing the input2d file formats for square vs. rectangular computational grids. Note that this is where the user specifies the size and resolution of the grid for the entire simulation.}
    \label{fig:input2d_SqToRec}
\end{figure}

The process of going from square to rectangular grids is only a few changes to the \textit{input2d} file, as illustrated in Figure \ref{fig:input2d_SqToRec}.

\textbf{$2^{nd}$ Order Time-Stepping IB Scheme} $ $\\

In essence this scheme incorporates the discretizations of the necessary operators in (\ref{Navier_Stokes} and (\ref{Incompressibility}) and uses a similar approach of the time-stepping scheme previously described in \cite{Peskin:1996,Battista:2015,BattistaIB2d:2016}. However, rather step one full unit forward in time, the previous method is used to step forward from $n\rightarrow n+\frac{1}{2}$. Next, a new scheme is introduced to move from $n\rightarrow n+1$ using the auxiliary variables found at time $n+\frac{1}{2}$. The motivation is to be able to use the Crank-Nicholson formalism, where the non-linear convective term is handled with a skew-symmetric finite difference approximation. Furthermore, this allows for a trapezoidal quadrature rule for the fiber model computations. This method was first introduced in \cite{Lai:2000}.

First, a preliminary step is used to find auxiliary variables $\textbf{u}^{n+1/2}$, $\tilde{p}^{n+1/2}$, $\textbf{X}^{n+1/2}$, as outlined below:

\begin{align}
\textbf{F}^{n}(s) &= \textbf{S}^{n}(\textbf{X}^{n}) \\ \nonumber \\
\textbf{f}^{n}(\textbf{x}) = \sum_{s} \textbf{F}^{n}&(s) \Delta_h(\textbf{x}-\textbf{X}^{n}(s)) \Delta s \\ \nonumber \\
\rho\Bigg( \frac{ \textbf{u}^{n+1/2}-\textbf{u}^{n} }{\Delta t/2} + \sum_{i=1}^2 u_{i}^n D_{i}^{\pm} \textbf{u}^{n} \Bigg) = &-\textbf{D}^0 \tilde{p}^{n+1/2} + \mu \sum_{i=1}^{2} D_{i}^{+} D_{i}^{-} \textbf{u}^{n+1/2} + \textbf{f}^n \\ \nonumber \\
\textbf{D}^{0}\cdot \textbf{u}&^{n+1/2} = 0 \\ \nonumber \\
\frac{ \textbf{X}^{n+1/2}(s) - X^{n}(s) }{\Delta t/2} = \sum_{x}&\textbf{u}^{n+1/2}(\textbf{x}) \Delta_h( \textbf{x} - \textbf{X}^{n}(s) ) h^2.
\end{align}

We note this preliminary step was the previous time-stepping method in \textit{IB2d} in \cite{Battista:2015,BattistaIB2d:2016}. Now to go from $n\rightarrow n+1$, for $\textbf{u}^{n+1/2}\rightarrow \textbf{u}^{n+1}$, $\tilde{p}^{n+1/2}\rightarrow p^{n+1/2}$, and $\textbf{X}^{n+1/2}\rightarrow \textbf{X}^{n}$, we do the following
\begin{align}
\textbf{F}^{n+1/2}(s) = &\textbf{S}^{n+1/2}(\textbf{X}^{n+1/2}) \\ \nonumber \\
\textbf{f}^{n+1/2}(\textbf{x}) = \sum_{s} \textbf{F}^{n+1/2}&(s) \Delta_h(\textbf{x}-\textbf{X}^{n+1/2}(s)) \Delta s \\ \nonumber \\
\nonumber \rho\Bigg( \frac{ \textbf{u}^{n+1}-\textbf{u}^{n} }{\Delta t} + \frac{1}{2}\sum_{i=1}^2 \Big( u_{i}^{n+1/2} D_{i}^{0} \textbf{u}^{n+1/2} &+ D_{i}^{0} \left(  u_{i}^{n+1/2} \textbf{u}^{n+1/2}    \right) \Big) \Bigg) = \\ -\textbf{D}^0 p^{n+1/2} &+ \frac{1}{2} \mu \sum_{i=1}^{2} D_{i}^{+} D_{i}^{-}\left( \textbf{u}^{n} + \textbf{u}^{n+1} \right)+ \textbf{f}^{n+1/2} \\ \nonumber \\
\textbf{D}^{0}\cdot \textbf{u}&^{n+1} = 0 \\ \nonumber \\
\frac{ \textbf{X}^{n+1}(s) - X^{n}(s) }{\Delta t} = \sum_{x}& \frac{ \textbf{u}^{n} + \textbf{u}^{n+1}  }{2} \Delta_h( \textbf{x} - \textbf{X}^{n+1/2}(s) ) h^2. 
\end{align}

We now define the the finite differencing operators. $\textbf{D}^0$ is the central differencing operator, defined as

\begin{equation}
\label{Discrete:CentralDiff_1} \textbf{D}^0 = \left( D_1^0, D_2^0 \right),
\end{equation}

with 

\begin{equation}
\label{Discrete:CentralDiff_2} \left(D_{\alpha}^0\phi\right)(\textbf{x}) = \frac{ \phi\big( \textbf{x} + \Delta \textbf{x} e_{\alpha} \big) - \phi\big( \textbf{x} - \Delta\textbf{x} e_{\alpha} \big) }{2\Delta x},
\end{equation}

where $\left(e_1,e_2\right)$ is the standard basis in $\mathbb{R}^2.$ The viscous term, given by $ \sum_{\alpha=1}^2 D_{\alpha}^{+} D_{\alpha}^{-} \textbf{u}$, is a difference approximation to the Laplacian, where the $D_{\alpha}^{\pm}$ operators are the forward and backward approximations to $\frac{ \partial}{\partial x_{\alpha} }.$ The upwind operators are defined as

\begin{equation}
\label{Discrete:Upwind} u_{\alpha}^{n} D_{\alpha}^{\pm} = \left\{ \begin{array}{c} 
u_{\alpha}^{n} D_{\alpha}^{-}  \ \ \ \ \ \ \ \ u_{\alpha}^{n} > 0, \\
u_{\alpha}^{n} D_{\alpha}^{+}  \ \ \ \ \ \ \ \ u_{\alpha}^{n} < 0,
\end{array} \right.
\end{equation}
with

\begin{align}
\label{Discrete:Forward} \left (D_{\alpha}^+\phi\right)(\textbf{x}) &= \frac{ \phi\big( \textbf{x} + \Delta \textbf{x} e_{\alpha} \big) - \phi\big( \textbf{x} \big) }{\Delta x} \\ \nonumber \\
\label{Discrete:Backward} \left (D_{\alpha}^{-}\phi\right)(\textbf{x}) &= \frac{ \phi\big( \textbf{x} \big) - \phi\big( \textbf{x} - \Delta\textbf{x} e_{\alpha} \big) }{\Delta x}. 
\end{align}

$ $\\


\textbf{Boussinesq Approximation} $ $\\

The Boussinesq approximation is incorporated into \textit{IB2d} to model fluctuations in the dynamics of both a concentration gradient (background field) and the momentum equation. The Boussinesq approximation can be thought of as an approximation to a variable density field, where the essence is that any differences in inertia are negligible, but gravity is strong enough to make the specific weight appreciably different between two fluids. 

In general the approximation ignores density differences except where they are multipled by a gravitational acceleration field, ${\bf{g}}$. It assumes that density variables have no effect on the fluid flow field, only that they give rise to bouyancy forces. By using the Boussinesq approximation, one bypasses the issue of having to solve the fully compressible Navier-Stokes equations for certain applications.

The extra forcing term on the incompressible Navier-Stokes equations (\ref{Navier_Stokes})-(\ref{Incompressibility}) are of the form
\begin{align}
{\bf{f}}_{Bouss} = \alpha_B \rho {\bf{g}} C,
\end{align}

where $\alpha_B$ is the expansion coefficient, e.g., thermal expansion, etc., $\rho$ is the density of the fluid, {\bf{g}} is the gravitational field, and $C$ is the background concentration. When implementating the Boussinesq approximation, the Navier-Stokes equations then take the form,

\begin{equation}
\label{Navier_Stokes_Bouss} \rho \left( \frac{\partial {\bf{u}}({\bf{x}},t) }{\partial t} + {\bf{u}}({\bf{x}},t)\cdot \nabla {\bf{u}}({\bf{x}},t) \right) = -\nabla p({\bf{x}},t) + \mu \Delta {\bf{u}}({\bf{x}},t) + {\bf{f}}({\bf{x}},t) +  \alpha_B \rho {\bf{g}} C \\
\end{equation}
\begin{equation}
\label{Incompressibility_Bouss} \nabla\cdot {\bf{u}}({\bf{x}},t) = 0.\\
\end{equation}

We note that these models have been incorporating into immersed boundary frameworks before, see \cite{Ghosh:2015,BGriffithIBAMR,Frisani:2012}. Examples using the Boussinesq approximation have been included below in Section \ref{Appendix_RT_Instability} and \ref{Appendix_Falling_Sphere} for the Rayleigh-Taylor Instability and Falling Spheres, respectively.

\section{Work Flow and Data Analysis}

While the workflow remains in large part the same as in \cite{Battista:2015, BattistaIB2d:2016}, there have been some subtle changes which allow for easier manipulation of examples and saving only desired simulation data.

%
%
%
%


%
%
%
%

\subsection{General Work Flow}

The typical work flow for using the \textit{IB2d} software is remains consistent between both MATLAB and Python implementations. Both implementations still have two subdirectories: an 'Examples' and 'IBM\textunderscore Blackbox' directory. The Examples subfolder contains all simulations that come with the software, including all necessary input files to run each simulation, e.g., the \textit{main2d.m} (or \textit{main2d.py}) and \textit{input2d} files, as well as all the input files associated for construction of that example's fiber model. The script for creating those fiber models is also included. The IBM\textunderscore Blackbox folder contains all scripts for performing the actual time-stepping routine in solving the FSI problem. The user does not have to modify any scripts in this subfolder, unless they wish to add to the already existing framework.\\

\begin{figure}[H]
    \centering
    \includegraphics[width=0.95\textwidth]{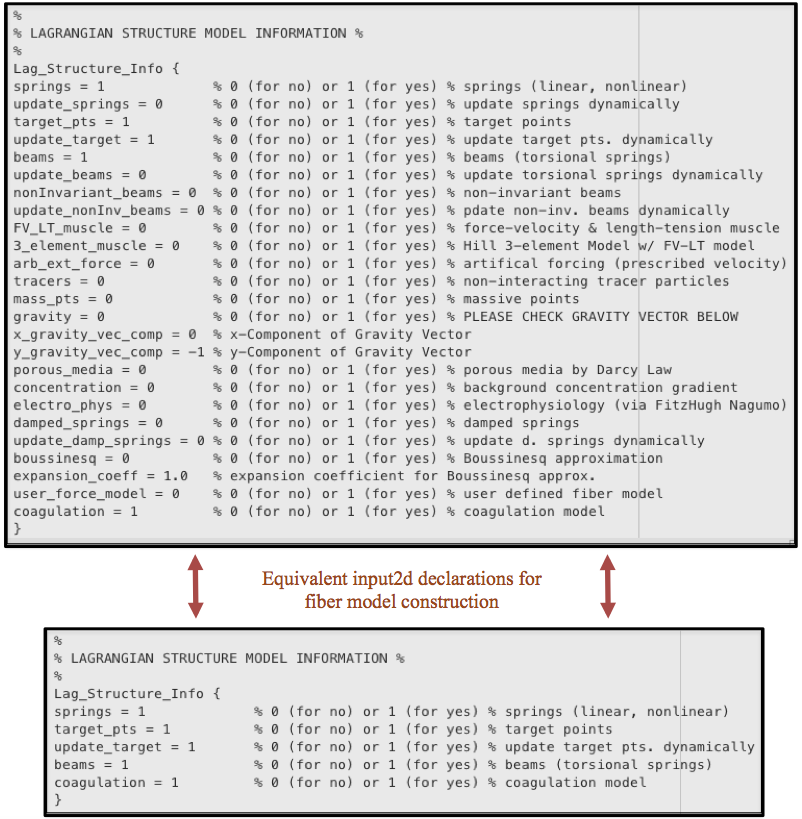}
    \caption{Two equivalent Lagrangian Structure Information selections in the \textit{input2d} files which declare equivalent fiber models that suffice to run a simulation.}
    \label{IB2d:input2d_fiber}
\end{figure}

There have been subtle changes to the \textit{input2d} file. Namely, a user only has to declare the fiber models they want to include in their example, e.g., if there are no non-invariant beam models in their model, they do not need to include a flag specifying that there are none. This is illustrated below in Figure \ref{IB2d:input2d_fiber}.  \\

\begin{figure}[H]
    \centering
    \includegraphics[width=0.85\textwidth]{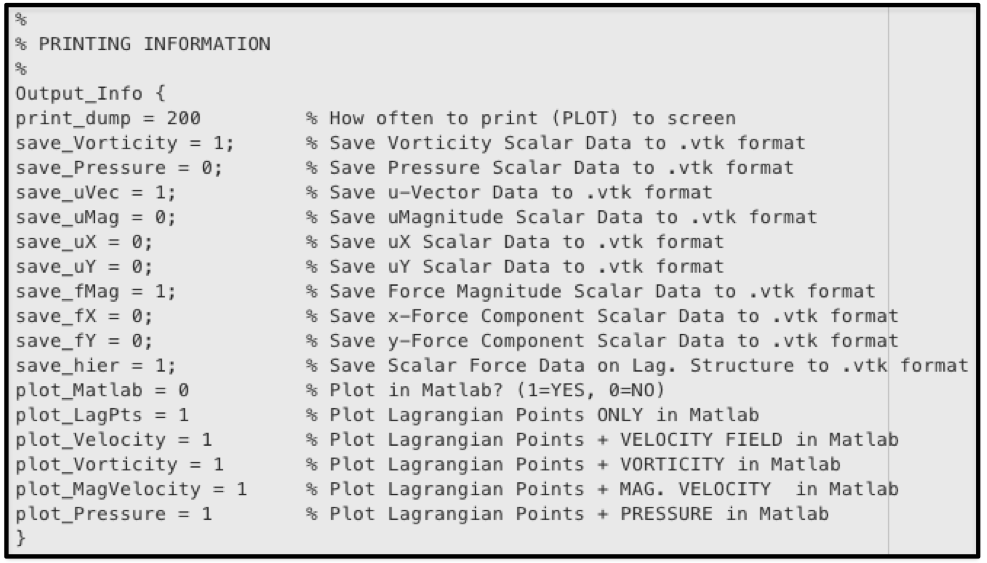}
    \caption{An example of Output Information flags in the \textit{input2d} file, such that as the simulation progresses the only data that will be saved is the scalar Vorticity data, vector velocity data, scalar Eulerian force magnitude data, and the force data on the Lagrangian Structure (``save\textunderscore hier"). Note that the Lagrangian $(x,y)$ positional data is automatically saved by default. Furthermore, the simulation will not plot information in MATLAB (or Python) as the simulation progresses, and it will save the data every $200$ time-steps (``print\textunderscore dump").}
    \label{IB2d:input2d_vtk}
\end{figure}

Furthermore, if data storage becomes an issue, a user can declare what data they wish to save to \textit{.vtk} file format for later analysis and/or visualization. An example is shown in Figure \ref{IB2d:input2d_vtk}, which illustrates the only simulation data that will be saved is the scalar vorticity data, vector velocity data, scalar Eulerian force magnitude data, and the force data on the Lagrangian Structure (``save\textunderscore hier"). Note that the Lagrangian $(x,y)$ positional data is automatically saved by default. Moreover, in this example the simulation will not dynamically plot any data in MATLAB (or Python) as the simulation progresses, and it will save the data every $200$ time-steps (given by the ``print\textunderscore dump" flag). Note that all the Eulerian data will be found in a \textit{viz\textunderscore IB2d} folder and all the force data on the Lagrangian structure will be found in the \textit{hier\textunderscore data\textunderscore IB2d} and is saved in the \textit{.vtk} format. This format can be visualized using Paraview \cite{Paraview:2005} or VisIt \cite{HPV:VisIt}.

Once the data is saved, the user can then  analyze the \textit{.vtk} data in the Data Analysis Toolbox, which is discussed in Section \ref{IB2d:Data_Analysis}. 

\begin{figure}[H]
    \centering
    \includegraphics[width=0.9\textwidth]{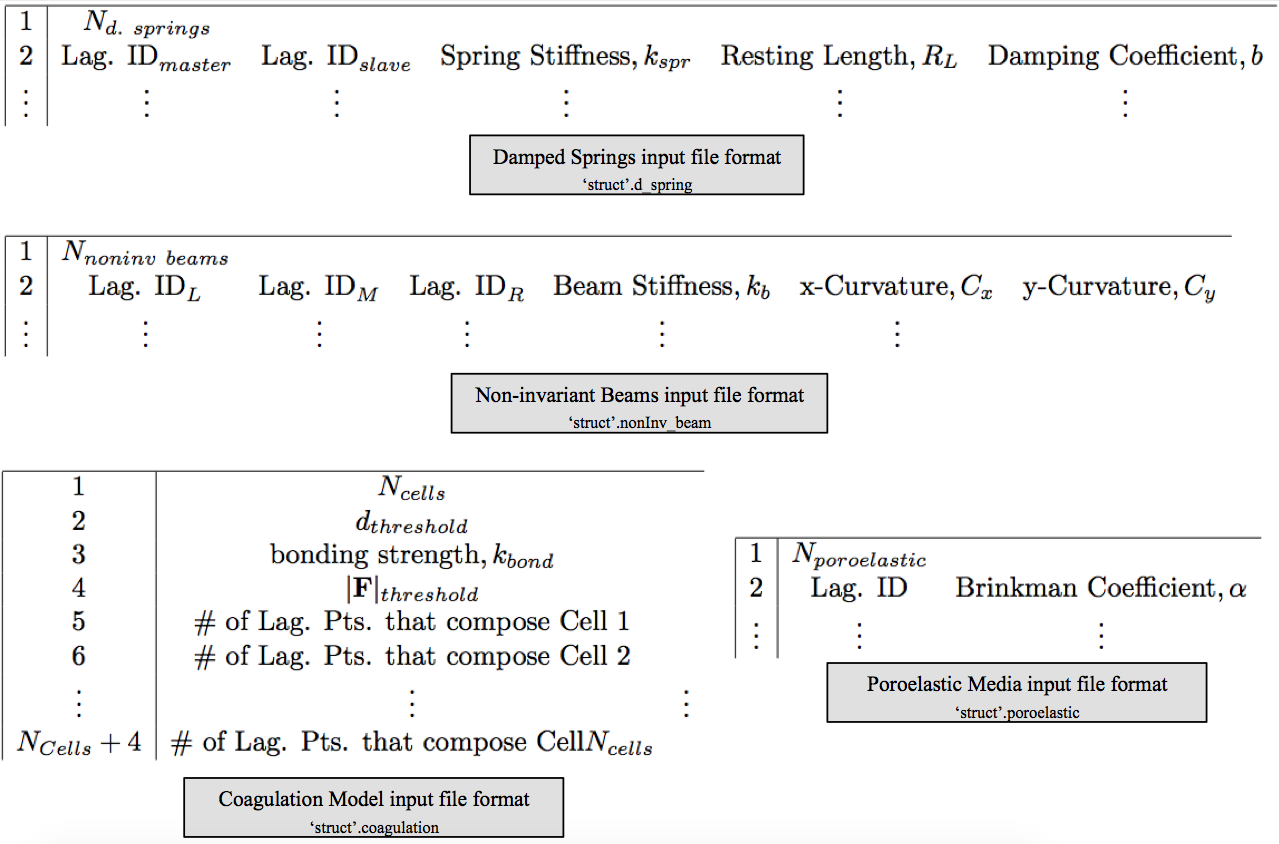}
    \caption{Input file formats for damped springs, non-invariant beams, poroelastic media, and coagulation/aggregation models.}
    \label{IB2d:Input_Formats}
\end{figure}

As with prior releases of \textit{IB2d}, each fiber model has an associated input file type. The fiber models that were previously implemented, e.g.
\begin{itemize}
\item Springs (linear, non-linear)
\item Beams (torsional springs)
\item Target Points
\item Massive Points
\item FV-LT Muscle
\item 3-Element Hill Model
\item Porous Media (via Darcy's Law)
\end{itemize}
have input file formats, as described in \cite{BattistaIB2d:2016}. If the immersed structure is designated as \textit{`struct'} in the \textit{input2d} file, then the fiber models introduced here, i.e., damped springs, non-invariant beams, poroelastic media, and coagulation/aggregation, have associated input formats as shown in Figure \ref{IB2d:Input_Formats}.

To run a simulation one only needs MATLAB or Python 3.5+ with the standard python libraries and NumPy, Numba and Matplotlib (if plotting is desired as the simulation progresses), all of which are present in a standard Anaconda Python 3.x distribution. VTK and pyFFTW are optional Python libraries that result in data write and computational speedups respectively. For further speedup of the data write process, one can also use Cython to compile included C libraries which will then automatically be recognized and used by the Python version of $IB2d$. 

In either version, the user then has only to enter an example folder (or create an example themselves) and run the \textit{main2d.m} (or \textit{main2d.py}) script. Currently there are over $60$ examples one can run upon downloading the software from \url{http://www.github.com/nickabattista/IB2d/}. After the simulation has completed, or surpassed the desired number of time-steps, one can visualize the Eulerian or Lagrangian data using the open source software of Paraview \cite{Paraview:2005} or VisIt \cite{HPV:VisIt} as mentioned before. While those open source visualization tools can offer some of their own data analysis features, \textit{IB2d} itself comes with a Data Analysis Toolbox for quantifying desired simulation data, as discussed in the proceeding section.

%
%
%
%

\subsection{\textit{IB2d} Data Analysis Package}

\label{IB2d:Data_Analysis}

\textit{IB2d} includes a data analysis package, which converts the data (\textit{.vtk}) files back into useful data structures in MATLAB or Python 3.5. Once imported, the data can then be manipulated appropriately. 

The data is imported using three different functions:
\begin{enumerate}
\item \textit{give\textunderscore Lag\textunderscore Positions()}: gives all the Lagrangian positions at a specific time-step
\item \textit{import\textunderscore Eulerian\textunderscore Data()}: gives all the Eulerian grid data at a specific time-step
\item \textit{import\textunderscore Lagrangian\textunderscore Force\textunderscore Data()}: gives the force data on the Lagrangian structure at a specific time-step
\end{enumerate}

\begin{figure}[H]
    \centering
    \includegraphics[width=0.975\textwidth]{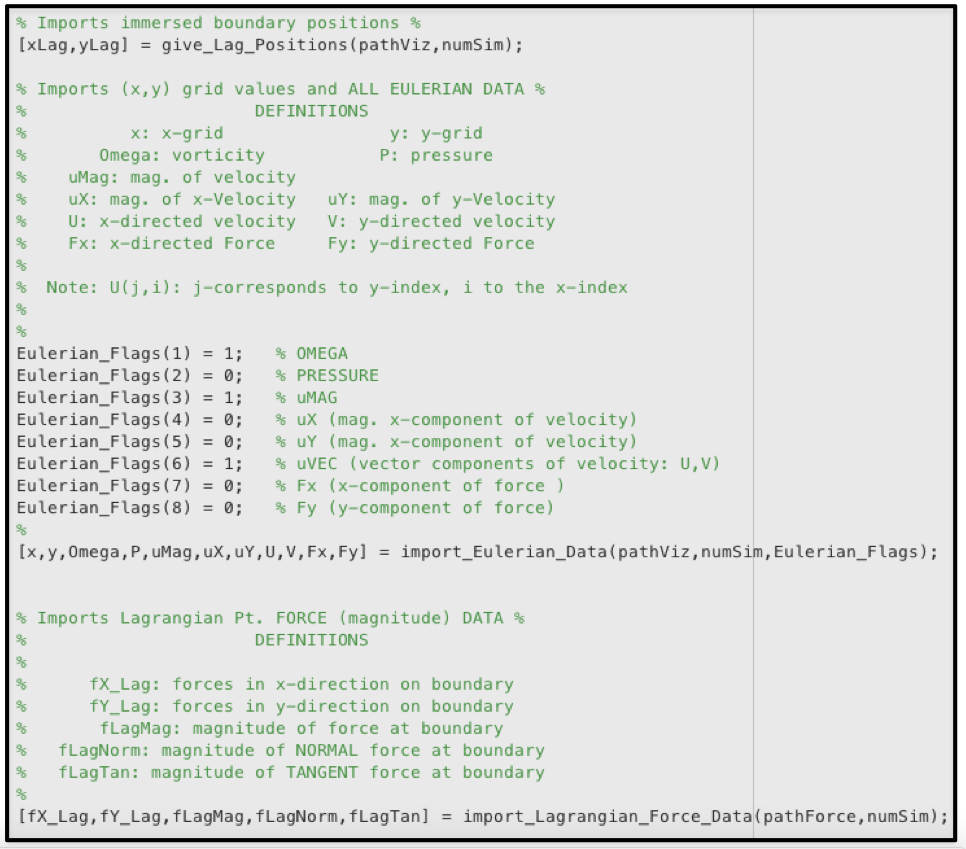}
    \caption{All the data (Lagrangian positions, Eulerian data, and Lagrangian force data) imported in the data analysis software.}
    \label{IB2d:Data_Analysis:Imports}
\end{figure}

Descriptions of all the data imported can be see in Figure \ref{IB2d:Data_Analysis:Imports}. Note that while reading in the Eulerian information, since not all the data is required to be printed in each simulation (see Figure \ref{IB2d:input2d_vtk}), one can choose what data to read in. In this example shown in Figure \ref{IB2d:Data_Analysis:Imports}, only the scalar vorticity, scalar magnitude of velocity, and vector velocity field are imported. Furthermore, the Lagrangian information imported from \textit{give\textunderscore Lag\textunderscore Positions()} will always be available since, by default, Lagrangian $(x,y)$ coordinates are saved. Moreover if one does not print the force data on the Lagrangian Structure, e.g. there is no \textit{hier\textunderscore data\textunderscore IB2D} folder because the \textit{print\textunderscore hier} flag in \textit{input2d} was set to zero, one can simply comment out the associated input command.

An example is contained within the code that analyzes data from a parabolic channel flow example that computes the magnitude of the velocity across multiple cross-sections of the channel. It is available in both the data analysis package for MATLAB and for Python 3.5. This simulation uses the following fiber model and functionality:

\begin{itemize}
    \item Target Points (fixed)
    \item Artificial Forcing on Fluid Grid
\end{itemize}

\begin{figure}[H]
    \centering
    \includegraphics[width=0.9\textwidth]{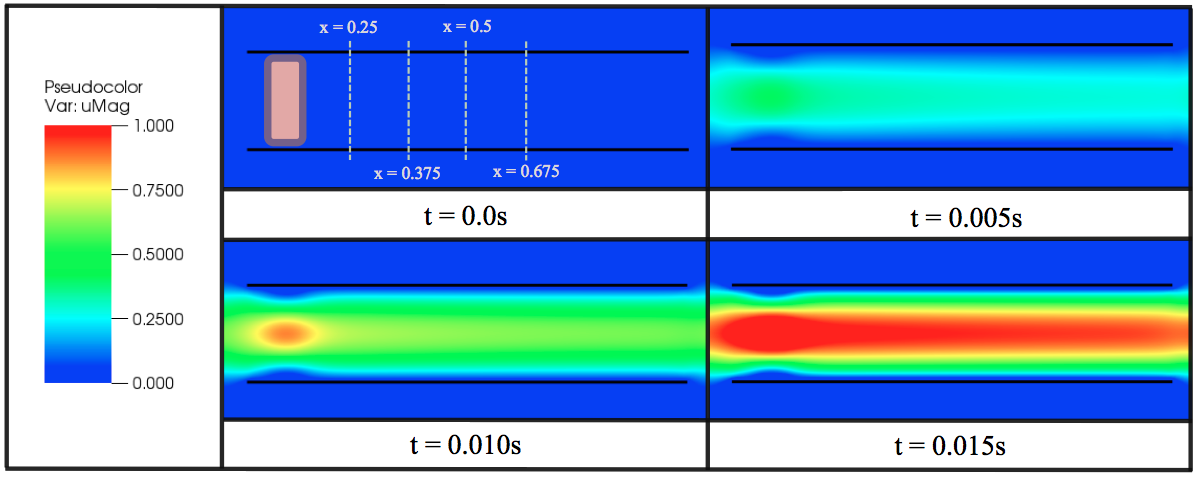}
    \caption{Simulation images taken from a channel with a parabolic flow condition at varying times. The parabolic flow is enforced by an external forcing condition on the Eulerian grid in the section outline in purple and shaded in red, while the vertical lines correspond to the cross-sections of the tube where the velocity data will be analyzed.}
    \label{IB2d:Data_Analysis:ChannelSims}
\end{figure}

\begin{figure}[H]
    \centering
    \includegraphics[width=0.95\textwidth]{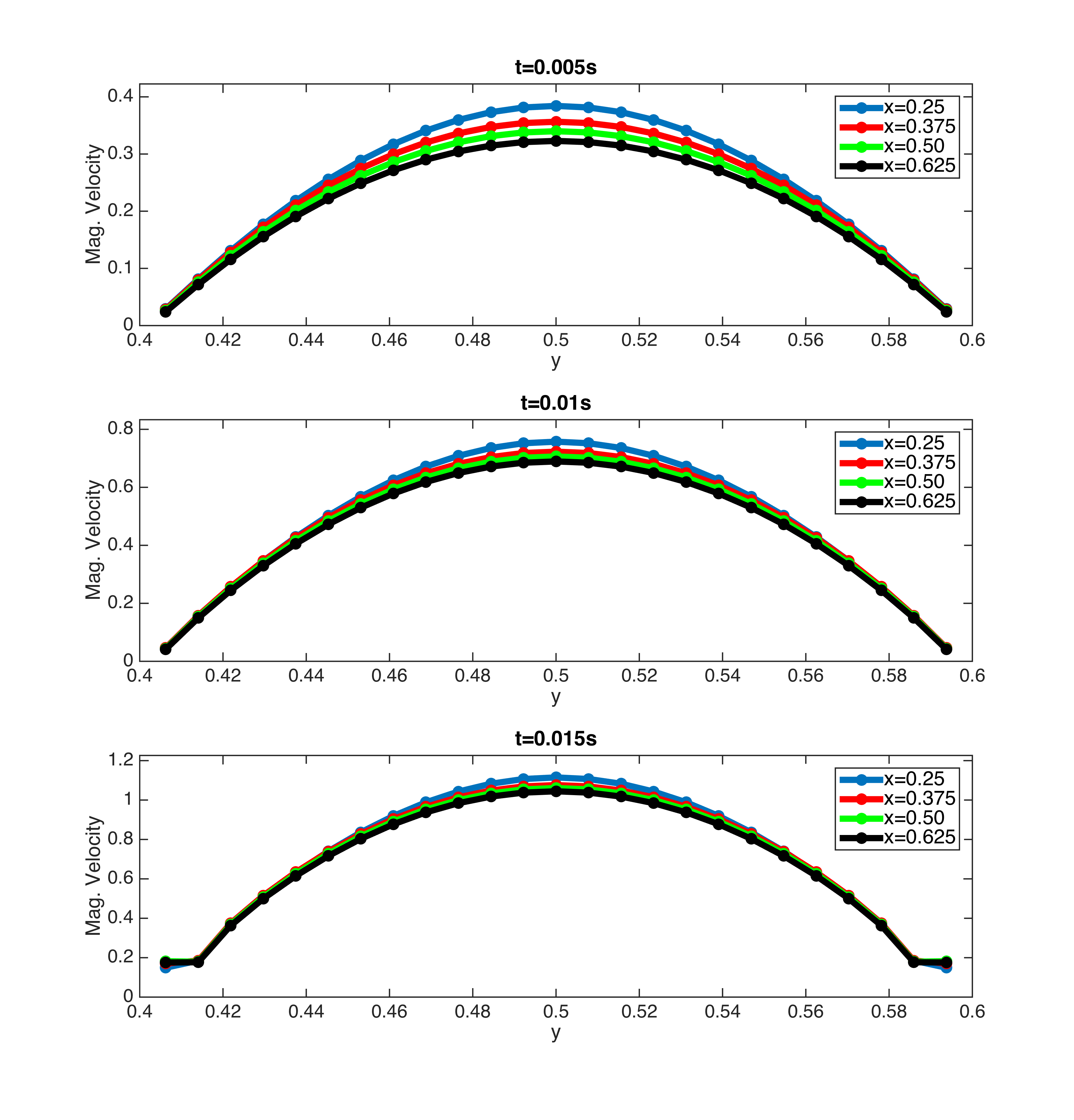}
    \caption{Data shown from three different time points during the simulation for velocities across four different cross-sections of the tube. As time increases, the velocity profile becomes more fully developed.}
    \label{IB2d:Data_Analysis:Channel}
\end{figure}

Simulation images are shown in Figure \ref{IB2d:Data_Analysis:ChannelSims}, which illustrate the magnitude of velocity in the channel, and data from the simulation is given in Figure \ref{IB2d:Data_Analysis:Channel}.

It is clear from Figure \ref{IB2d:Data_Analysis:Channel} that as the simulation progresses the velocity profile within a cross-section of the tube fully develops. The data plotted was taken along the dashed-vertical lines from Figure \ref{IB2d:Data_Analysis:ChannelSims}.

%
%
%
%
%
%

%
%
%
%
%
%

\section{Examples illustrating new functionality}
\label{IB2d_Examples}

In this section we will present some examples which show some of the software's new functionality. The software currently contains over $50$ built-in examples, some of which come from previous IB papers in the literature. We will show the following examples:

\begin{itemize}
    \item Tethered Ball in Channel
    \item Anguilliform Swimmer
    \item Rayleigh-Taylor Instability
    \item Falling Spheres
    \item Seagrass in Oscillatory Flow
    \item Stirring Cells with Coagulation
\end{itemize}

$ $\\


\subsection{Tethered Ball in Channel}
\label{IB2d:channel_ball}
$ $\\

\begin{figure}[H]
    \centering
    \includegraphics[width=0.975\textwidth]{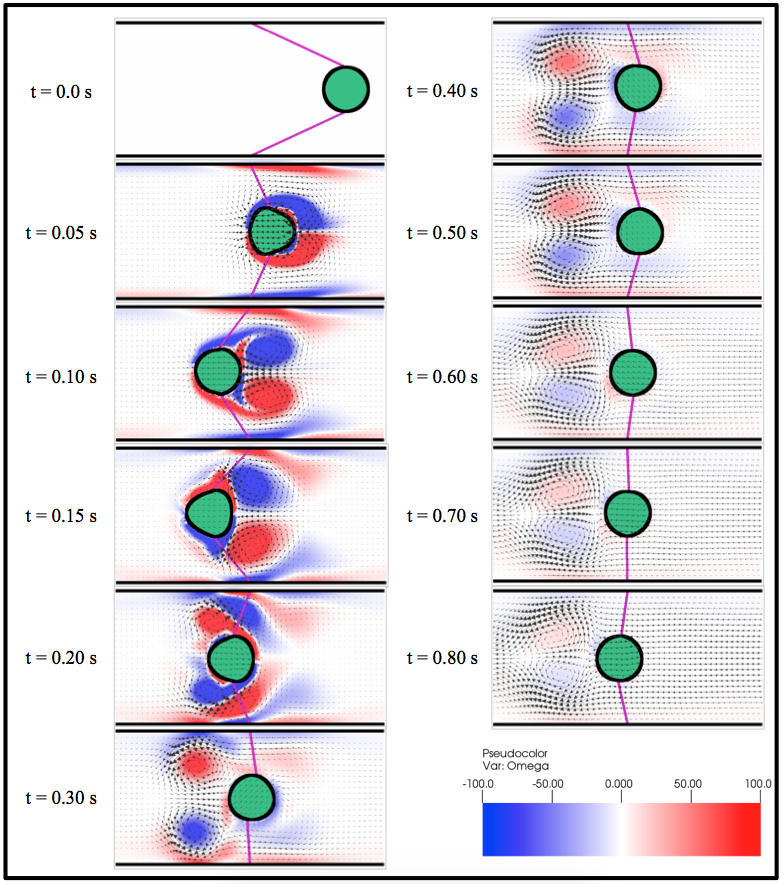}
    \caption{A tethered elastic ball is stretched in a channel and released. It is tethered to the channel walls using damped springs. The colormap shows vorticity and the background vector field is the fluid's velocity.}
    \label{Example:Channel_Ball}
\end{figure}

This example uses damped linear springs to tether an elastic ball to the channel walls. The ball is composed of linear springs and torsional springs along adjacent Lagrangian points and linear springs connecting a Lagrangian point on the ball to its counterpart on the opposite side of the ball. The channel is composed of target points that are being held nearly rigid. The following fiber models were used:
\begin{itemize}
    \item Linear Springs
    \item Damped linear springs
    \item Torsional Springs 
    \item Target Points
\end{itemize}

The simulation begins with the damped springs stretched from their equilibrium position, where the preferred position is with the tethering damped springs configured vertically in the channel. The ball is released immediately upon the start of the simulation and oscillates back and forth, before eventually settling down in its preferred configuration. An example simulation is shown in Figure \ref{Example:Channel_Ball}.

\begin{figure}[H]
    \centering
    \includegraphics[width=0.675\textwidth]{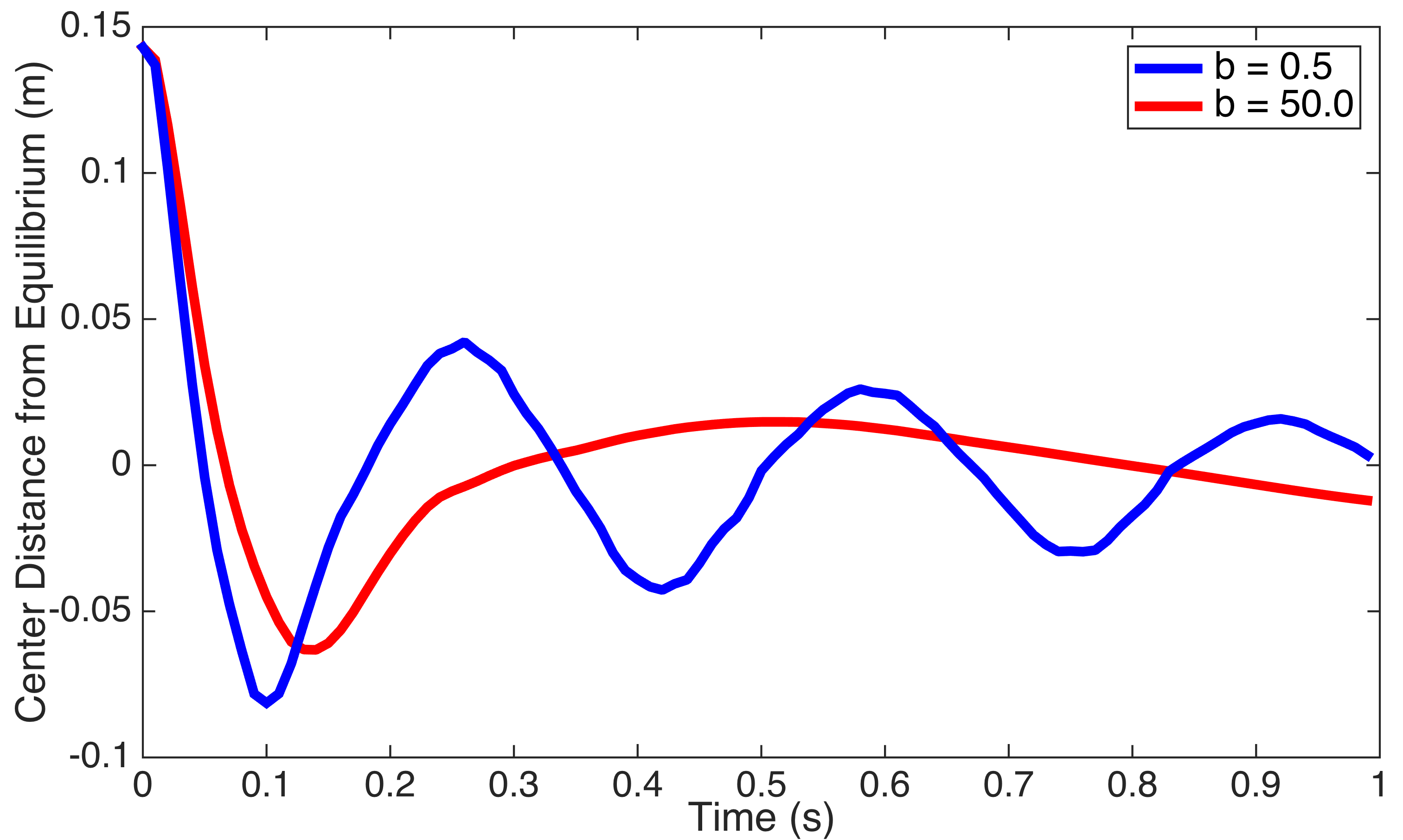}
    \caption{A comparison of two simulations with different damping coefficients. The case with less damping shows oscillatory behavior, while the case with more damping almost appears critically damped.}
    \label{Example:Ball_Data}
\end{figure}

Using the Data Analysis package in \textit{IB2d}, as described in Section \ref{IB2d:Data_Analysis}, a comparison between two simulations with differing damping strengths is shown. These results are illustrated in Figure \ref{Example:Ball_Data}. The case in blue ($b=0.5$) has a lower damping coefficient than the case in red ($b=50.0$). In the lower damping case, there are more oscillations before equilibrium is achieved, while in the higher damping case, the system appears almost critically damped, and almost immediately returns to its preferred position.\\

\subsection{Idealized Anguilliform Swimmer}
\label{IB2d:beam_swimmer}
$ $\\

\begin{figure}[H]
    \centering
    \includegraphics[width=0.995\textwidth]{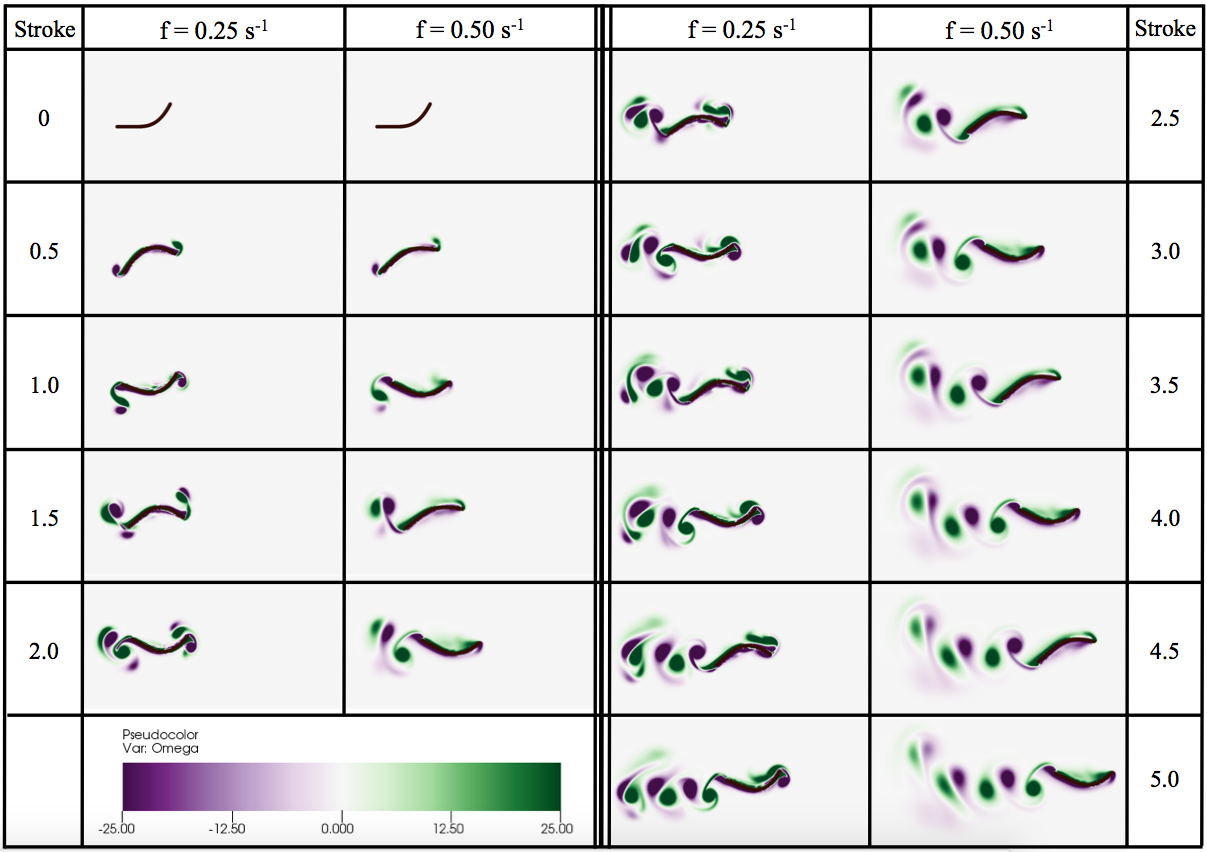}
    \caption{A comparison of two idealized anguilliform swimmers moving forward due to continually changes in the preferred curvature of the configuration. One has a stroke frequency of $f=0.25s^{-1}$ and the other, $f=0.5s^{-1}$. The colormap illustrates vorticity.}
    \label{Example:Beam_Swimmer}
\end{figure}

This example uses non-invariant beams and dynamically updates the preferred beam curvature through the `\emph{update$\_$nonInv$\_$Beams}' script to move forward. The model also uses linear springs to connect successive Lagrangian points and all successive Lagrangian points are connected by non-invariant beams. The following fiber models were used:
\begin{itemize}
    \item Linear Springs
    \item Non-Invariant Beams
\end{itemize}

The motion is completely induced by changing the preferred curvature.  Within the `\emph{update$\_$nonInv$\_$Beams}', the curvature is changed by interpolating through two different configurative phases of the swimmer, more specifically their associated curvatures of each phase. The swimming motion is illustrating in Figure \ref{Example:Beam_Swimmer}. Those phases are shown below,

\begin{figure}[H]
    \centering
    \includegraphics[width=0.8\textwidth]{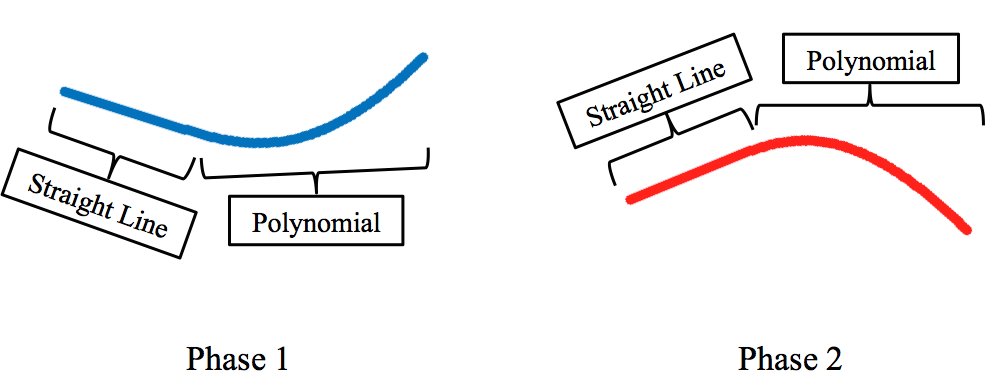}
    \caption{The two phases, in which, the preferred curvature was interpolated between to cause forward swimming.}
    \label{Example:Beam_Swimmer_Phases}
\end{figure}

Using the Data Analysis package in \textit{IB2d}, which is described in Section \ref{IB2d:Data_Analysis}, a comparison of distances swam by each swimmer as a function of the number of strokes is shown. The swimmer with the slower stroke frequency performs better.\\

\begin{figure}[H]
    \centering
    \includegraphics[width=0.75\textwidth]{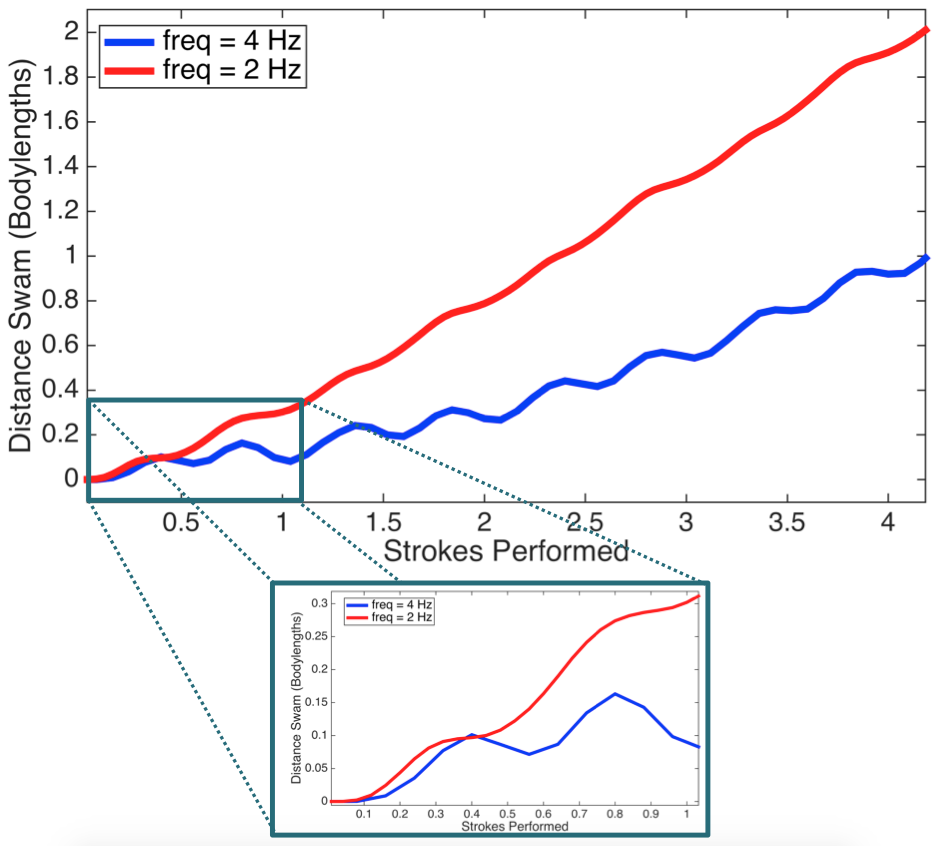}
    \caption{A comparison of the distances swam by both swimmers as a function of the number of strokes.}
    \label{Example:Beam_Swimmer_Performance}
\end{figure}


\subsection{Rayleigh-Taylor Instability}
\label{Appendix_RT_Instability}

\begin{figure}[H]
    \centering
    \includegraphics[width=0.8\textwidth]{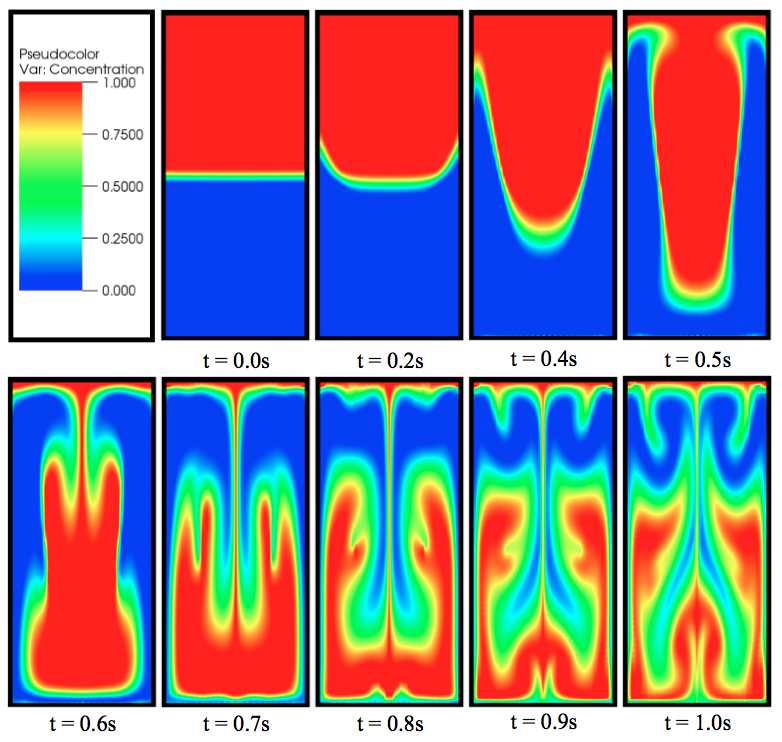}
    \caption{Simulation of the Rayleigh-Taylor Instability using the Boussinesq Approximation. A heavier fluid (red) sits above a lighter fluid (blue). Note a more sophisticated advection-diffusion solver will give rise to higher resolution of the instability fronts.}
    \label{Example:Rayleigh_Taylor}
\end{figure}

This example uses the Boussinesq Approximation to model the Rayleigh-Taylor Instability. The instability manifests itself at the interface between two fluids of different densities when the lighter fluid begins pushing the heavier fluid. The model itself only uses target points to bounday the fluids in a rectangular domain, concentration gradients, and the Boussinesq approximation frameworks. Hence the fiber models and functionality used are:

\begin{itemize}
    \item Target Points (fixed)
    \item Background Concentration (advection-diffusion)
    \item Boussinesq Approximation (with gravity flag)
\end{itemize}

The simulation begins when a heavier fluid (red) is placed over a lighter fluid (blue) with a linear change in concentration at the interface. The lighter fluid begins pushing itself upwards while the heavier fluid falls downward, resulting in the instability. The dynamics can be seen in Figure \ref{Example:Rayleigh_Taylor}. Note a more sophisticated advection-diffusion solver will give rise to higher resolution of the instability fronts. Operator splitting methods \cite{Chertock:2012} and flux limiters \cite{LeVeque:2002} are currently being implemented.


\subsection{Falling Sphere with Boussinesq Approximation}
\label{Appendix_Falling_Sphere}

\begin{figure}[H]
    \centering
    \includegraphics[width=0.975\textwidth]{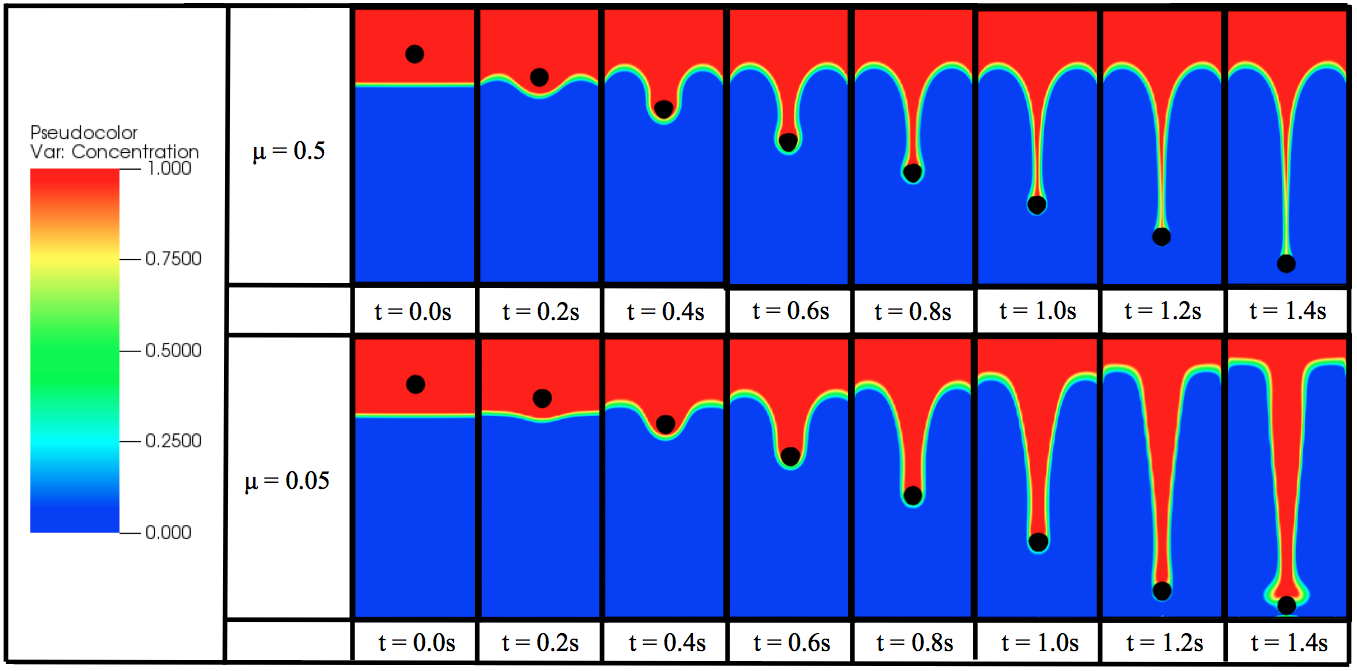}
    \caption{Simulation of a falling sphere through a fluid with a background salinity stratification. A spherical mass is released in the lighter salinity background (red) that sits above a heavier background salinity (blue) and the mass falls due to gravity. }
    \label{Example:Falling_Sphere_Concentration}
\end{figure}

This example uses massive points to model a sphere that is released in a lighter salinity background and then falls into the heavier salinity portion. The sphere is composed of massive points along the boundary, springs connecting adjacent Lagrangian points and the associated Lagrangian point across the sphere, and beams around adjacent points on the sphere. The domain itself is composed of target points. Hence the fiber models and functionality used are: 

\begin{itemize}
    \item Springs (linear)
    \item Beams (torsional springs)
    \item Target Points (fixed)
    \item Massive Points (w/ gravity)
    \item Background Concentration (advection-diffusion)
    \item Boussinesq Approximation (with gravity flag)
\end{itemize}

When the sphere begins falling, it entrains some of the lighter salinity concentration around carrying the lighter fluid downward. Two cases are compared corresponding to different ambient fluid viscosities. The sphere in the less viscous fluid falls faster than the higher viscosity case, and it also entrains more of the lighter salinity concentration. Snapshots of the simulation can be seen in Figure \ref{Example:Falling_Sphere_Concentration}.\\


\subsection{Seagrass in Oscillatory Flow}
\label{Section_Seagrass}

\begin{figure}[H]
    \centering
    \includegraphics[width=0.975\textwidth]{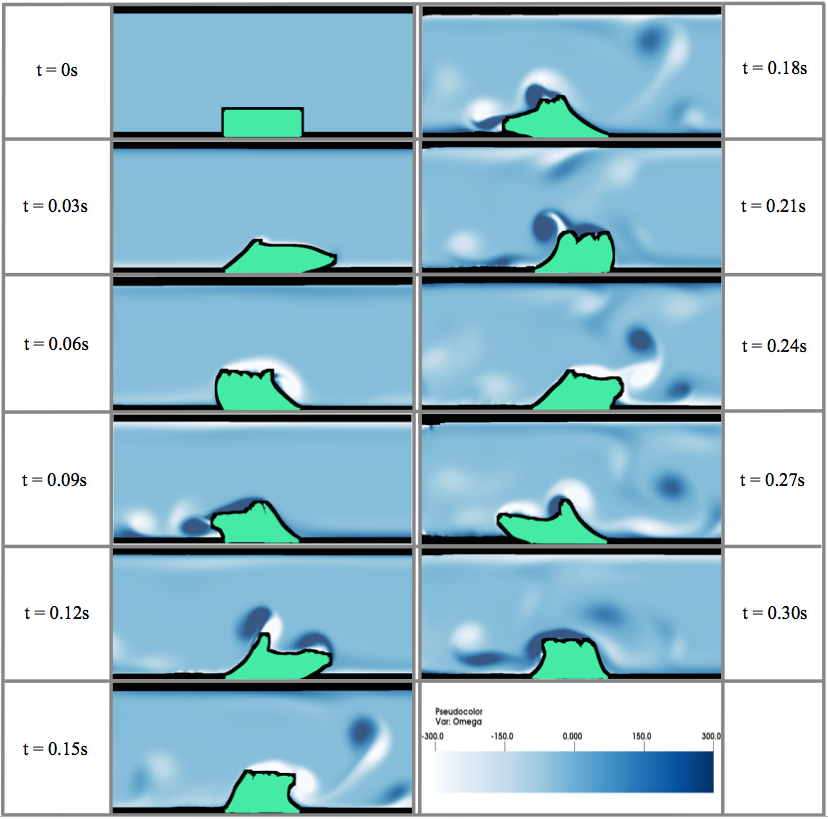}
    \caption{Simulation of seagrass in oscillatory flow. The seagrass (in green) bends towards the right and the left depending on the magnitude and direction of flow. The background colormap is vorticity.}
    \label{Example:Seagrass}
\end{figure}

This example uses poroelastic media to model seagrass. Each Lagrangian point that composes the seagrass is tethered to its neighboring points using linear springs, as in Figure \ref{FiberModel:Poroelastic}. A background oscillatory flow is initiated, which causes the seagrass to wave towards the right and left depending on the magnitude and direction of flow. The channel domain is composed of target points. Hence the fiber models and functionality used are: 

\begin{itemize}
    \item Springs (linear)
    \item Target Points (fixed)
    \item Poroelastic Media
    \item Artificial Forcing on the Fluid Grid (to induce oscillatory flow)
\end{itemize}

As the simulation begins, the flow first moves towards the right side of the domain and in response, the seagrass deforms toward the right. As the flow changes direction, the seagrass begins to return towards its equilibrium (original) position, and a large starting vortex is formed on the right side. As the flow continues and changes direction, this pattern of a starting vortex leading off the highest deformed side of the seagrass remains consistent. The swaying motion of the seagrass enhances mixing along the seafloor and above the seagrass itself, as seen in Figure \ref{Example:Seagrass}.\\


\subsection{Stirring with Coagulation}
\label{Section_Coagulation}

\begin{figure}[H]
    \centering
    \includegraphics[width=0.975\textwidth]{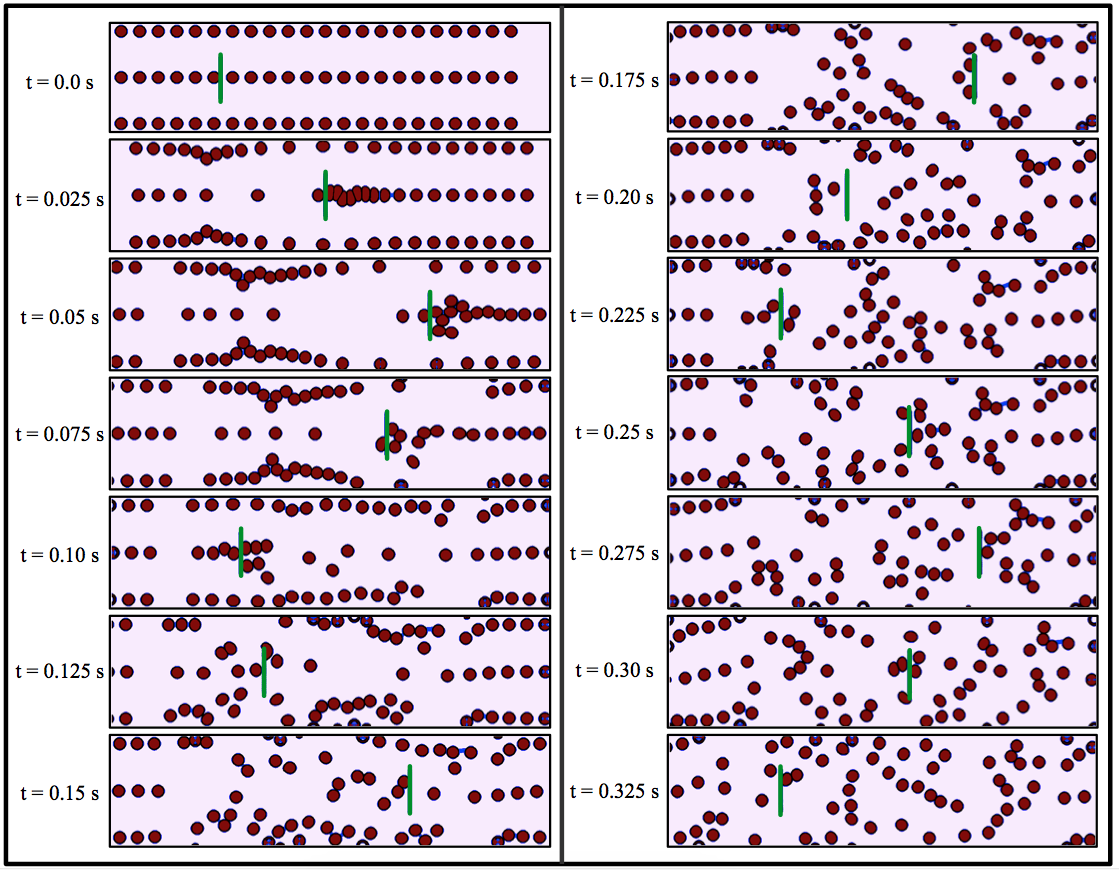}
    \caption{Simulation of cells being stirred in a rectangular box. The cells aggregate when they come close enough and form a bond between the two. The bond is fractured if the force exceeds the fracture force threshold.}
    \label{Example:Coagulation}
\end{figure}

This example uses coagulation and aggregation models to simulate cells being stirred. Each cell is composed of Lagrangian points around its circumference with linear springs adjoining adjacent nodes. There are also torsional springs along adjacent Lagrangian points of each cell. Moreover, there are linear springs tethering each Lagrangian points its opposite on the other side of the cell. The stirring apparatus is composed entirely of target points. The cells form a bond (linear spring) with another cell if the cells come into close enough contact, e.g. a distance less than the bonding threshold. The fiber models and functionality used are:

\begin{itemize}
    \item Springs (linear)
    \item Torsional Springs
    \item Target Points (prescribed motion)
    \item Coagulation Models
\end{itemize}

\begin{figure}[H]
    \centering
    \includegraphics[width=0.975\textwidth]{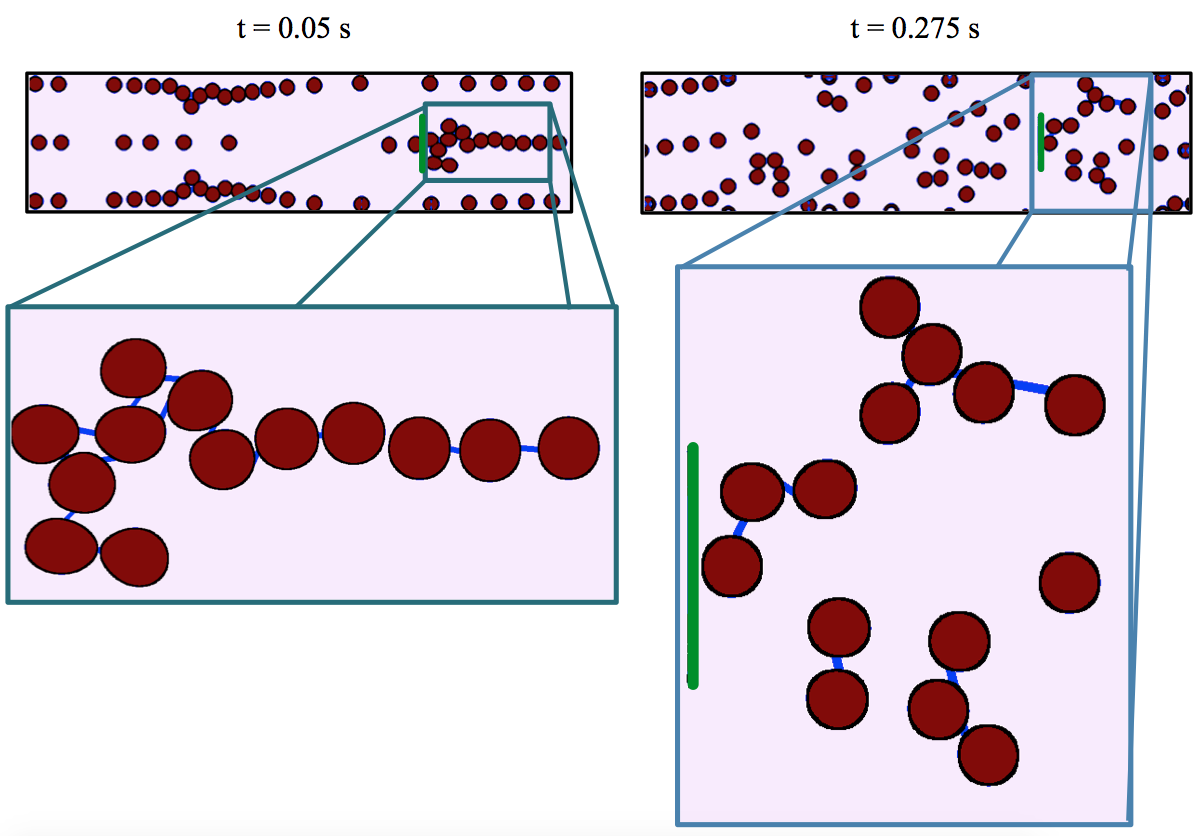}
    \caption{Illustration of bond formation at two different times during the simulation, $t=0.05s$ (left) and $t=0.275s$ (right).}
    \label{Example:Coag2}
\end{figure}

When the simulation begins, the stirrer moves towards the right of the domain. As it moves right, it initiates the fluid motion that pushes the cells around. The stirrer changes direction and moves back left causing the fluid to change direction. As the fluid changes its direction, the cells continually are being pushed and pulled throughout the domain, either by the fluid or by other cells if a bond has formed. If the forces arising from the bond formed between two cells is greater than the fracture threshold, the bond is destroyed. Snapshots of the dynamics can be seen in Figure \ref{Example:Coagulation}.  An illustration of bond formation can be seen in Figure \ref{Example:Coag2}, which shows the bonds that have formed between cells in a subset of the domain at two different times in the simulation.

%
%

\section{Discussion and Conclusion}

\textit{IB2d} is immersed boundary software with full implementations in both MATLAB and Python 3.5 that has been enhanced since its original release \cite{Battista:2015,BattistaIB2d:2016}. It offers robust construction of the immersed structure's material properties through a vast array of fiber model options. It also includes functionality for concentration gradients, artificial forcing to prescribe flow profiles, muscle mechanics, poroelastic and porous structures, electrophysiology, and slight fluid density variations via the Boussinesq approximation. Furthermore, having been specifically written in high-level programming languages, it allows for easier readability of the scripts at the heart of the immersed boundary method and an accelerated rate of modification for adding functionality such as other fluid solvers. 

Although high-level programming languages come with a few drawbacks, such as slower computational speeds compared to lower-level languages, one can still use \textit{IB2d} for high resolution applications on rectangular grids. Due to computational costs, for high resolution simulations on square grids, e.g. grids more highly resolved than $512\times512$, we suggest moving to Boyce Griffith's open source IBAMR package \cite{IBAMR}. 

IBAMR is an adaptive and parallelized implementation of $IB$ written in $C++$ including functionality for a hybrid finite element-finite difference implementation of $IB$ \cite{Griffith:IBFE}. It is dependent on many open source libraries, including PETSc \cite{PETSc}, SAMRAI \cite{SAMRAI}, libMesh \cite{libMesh}, and OpenMPI \cite{openMPI} which makes it very efficient to run, but at the cost of a nontrivial installation process and steep learning curve for anyone inexperienced at high performance computing. Furthermore, without computer clusters available, IBAMR cannot run at its full potential.

Moreover, \textit{IB2d} was strictly designed for $2D$ applications. While full $3D$ simulations are often desired, some applications may only require fluids with two-dimensions \cite{Tytell:2010,Crowl:2011,Zhang:2014,Lewis:2015,Waldrop:2015}. \textit{IB2d} was written in $2D$ to make it more readable and to lend itself for easier modification, particularly as a first step in trying to implement a new model. If $3D$ applications are required, we suggest moving to IBAMR. Note that the format of \textit{IB2d} was designed to mirror the input file formats used in IBAMR, and as such \textit{IB2d} can used as a stepping stone to using IBAMR; however, \textit{IB2d} currently offers more fiber models than IBAMR. Moreover, neither \textit{IB2d} nor IBAMR include functionality for compressible fluids, non-Newtonian fluids, or variable density fluid applications at this time, but future implementations of \textit{IB2d} may incorporate them. 

Similarly, neither \textit{IB2d} nor IBAMR include a turbulence model for large $Re$ applications at this time. For these applications, one may be interested in OpenFOAM by OpenCFD LTD \cite{openFOAM:openFOAM}. OpenFOAM is open source software capable of FSI applications, written in $C++$; however, to add additional numerical schemes is nontrivial and involves a steep learning curve. Commercial software such as COMSOL \cite{COMSOL:2011} and ANSYS Fluent \cite{Fluent:2016} can model FSI, but licenses for them are expensive and once again, it is rather nontrivial to implement other numerical methods and models in their frameworks.

While Peskin's immersed boundary method offers an elegant framework for modeling fluid-structure interactions, there exist other methods for FSI, such as level set methods \cite{Sethian:1999,Osher:2002}, the blob projection method \cite{Cortez:2000}, sharp interface methods \cite{Ubbink:1999,Udaykumar:2001}, and immersed interface methods \cite{Lee:2003,Li:2003}. These methods all offer higher resolution near the fluid-structure interface; however, most are limited to thin structures and rigid bodies. Furthermore, no open source implementations are known at this time, requiring a large entry time for research. For further information and a broader perspective on immersed boundary methods see \cite{Mittal:2005}.

\textit{IB2d} is an ideal environment for entry into FSI. Not only does it offer a plethora of fiber models, functionality, and $50+$ examples illustrating the breadth of the software, but being written in high-level languages, it allows for fast implementations of new fiber models and constitutive models, fluid solvers, and other functionality in an $IB$ framework. The software is available at \url{http://www.github.com/nickabattista/IB2d}.


%
%

\section{Acknowledgements}
The authors would like to thank Charles Peskin for the development of immersed boundary method and Boyce Griffith for IBAMR, to which many of the input files structures of \textit{IB2d} are based. We would also like to thank Austin Baird, Christina Battista, Robert Booth, Namdi Brandon, Kathleen Carroll, Christina Hamlet, Alexander Hoover, Shannon Jones, Andrea Lane, Jae Ho Lee, Virginia Pasour, Julia Samson, Arvind Santhanakrishnan, Michael Senter, Anne Talkington, and Ben Vadala-Roth, Lindsay Waldrop for comments on the design of the software and suggestions for examples. This project was funded by NSF DMS CAREER \#1151478, NSF CBET \#1511427, NSF DMS \#1151478, NSF POLS \#1505061 awarded to L.A.M. Funding for N.A.B. was partially funded from an National Institutes of Health T32 grant [HL069768-14; PI, Christopher Mack] and partially by the UNC Dissertation Completion Fellowship.

%
%


%
%


\bibliographystyle{elsarticle-num}

\bibliography{heart}

\begin{thebibliography}{10}
\expandafter\ifx\csname url\endcsname\relax
  \def\url#1{\texttt{#1}}\fi
\expandafter\ifx\csname urlprefix\endcsname\relax\def\urlprefix{URL }\fi
\expandafter\ifx\csname href\endcsname\relax
  \def\href#1#2{#2} \def\path#1{#1}\fi

\bibitem{BattistaIB2d:2016}
N.~A. Battista, W.~C. Strickland, L.~A. Miller, Ib2d: a python and matlab
  implementation of the immersed boundary method, Bioinspir. Biomim. 12(3)
  (2017) 036003.

\bibitem{Bathe:2008}
K.~J. Bathe, Fluid-structure interactions, Mechanical Engineering April (2008)
  67--68.

\bibitem{Bak:2013}
S.~Bak, J.~Yoo, C.~{Yong Song}, Fluid-structure interaction analysis of
  deformation of sail of 30-foot yacht, Int. J. of Naval Arch. and Ocean Eng.
  5(2) (2013) 263--276.

\bibitem{Rao:2015}
K.~{Srinivasa Rao}, K.~{Girija Sravani and G.Yugandhar}, G.~{Venkateswara Rao},
  V.~N. Mani, Design and analysis of fluid structure interaction in a
  horizontal micro channel, Procedia Materials Science 10 (2015) 768--788.

\bibitem{Vanderhoydonck:2016}
B.~Vanderhoydonck, G.~Santo, J.~Vierendeels, J.~Degroote, Optimization of a
  human-powered aircraft using fluid-structure interaction simulations,
  Aerospace 3 (2016) 26.

\bibitem{Fauci:1988}
L.~Fauci, C.~Peskin, A computational model of aquatic animal locomotion, J.
  Comput. Phys. 77 (1988) 85--108.

\bibitem{Hershlag:2011}
G.~Hershlag, L.~A. Miller, Reynolds number limits for jet propulsion: a
  numerical study of simplified jellyfish, J. Theor. Biol. 285 (2011) 84--95.

\bibitem{Hoover:2015}
A.~P. Hoover, L.~A. Miller, A numerical study of the benefits of driving
  jellyfish bells at their natural frequency, J. Theor. Biol. 374 (2015)
  13--25.

\bibitem{Miller:2004}
L.~A. Miller, C.~S. Peskin, When vortices stick: an aerodynamic transition in
  tiny insect flight, J. Exp. Biol. 207 (2004) 3073‚Äì3088.

\bibitem{Ruck:2010}
S.~Ruck, H.~Oertel, Fluid-structure interaction simulation of an avian flight
  model, J. Exp. Biol. 213~(24) (2010) 4180--4192.

\bibitem{SJones:2015}
S.~K. Jones, R.~Laurenza, T.~L. Hedrick, B.~E. Griffith, L.~A. Miller, Lift-
  vs. drag-based for vertical force production in the smallest flying insects,
  J. Theor. Biol. 384 (2015) 105--120.

\bibitem{Wolters:2005}
B.~Wolters, M.~Rutten, G.~Schurink, U.~Kose, J.~{de Hart}, F.~{van de Vosse}, A
  patient-specific computational model of fluid-structure interaction in
  abdominal aortic aneurysms, Med. Eng. and Phys. 27(10) (2005) 871--883.

\bibitem{Torii:2008}
R.~Torii, M.~Oshima, T.~Kobayashi, K.~Takagi, T.~E. Tezduyar,
  Fluid‚Äìstructure interaction modeling of a patient-specific cerebral
  aneurysm: influence of structural modeling, Comp. Mech. 43 (2008) 151.

\bibitem{Takizawa:2011}
K.~Takizawa, C.~Moorman, S.~Wright, J.~Purdue, T.~McPhail, P.~R. Chen,
  J.~Warren, T.~E. Tezduyar, Patient-specific arterial fluid‚Äìstructure
  interaction modeling of cerebral aneurysms, Int. J. Num. Meth. Fluids
  65~(1-3) (2011) 308--323.

\bibitem{Yang:2007}
C.~Yang, D.~Tang, I.~Haber, T.~Geva, P.~J. {del Nido}, In vivo mri-based 3d fsi
  rv/lv models for human right ventricle and patch design for potential
  computer-aided surgery optimization, Computers and Structures 85 (2007)
  988--997.

\bibitem{Carson:2017}
E.~Carson, C.~Cobelli, Modeling Methodology for Physiology and Medicine,
  Elsevier Science, Amsterdam, Netherlands, 2017.

\bibitem{Santhanakrishnan:2009}
A.~Santhanakrishnan, N.~Nguyen, J.~Cox, L.~A. Miller, Flow within models of the
  vertebrate embryonic heart, J. Theor. Biol. 259 (2009) 449--461.

\bibitem{Miller:2011}
L.~A. Miller, Fluid dynamics of ventricular filling in the embryonic heart,
  Cell Biochem. Biophys. 61 (2011) 33--45.

\bibitem{Battista:2016a}
N.~A. Battista, A.~N. Lane, L.~A. Miller, On the dynamic suction pumping of
  blood cells in tubular hearts, \textit{in press} Association for Women in
  Mathematics Series: Research Collaborations in Mathematical Biology.

\bibitem{Battista:2016b}
N.~A. Battista, A.~N. Lane, J.~Liu, L.~A. Miller, Fluid dynamics of heart
  development: Effects of trabeculae and hematocrit, arXiv:
  https://arxiv.org/abs/1610.07510.

\bibitem{Peskin:1972}
C.~Peskin, Flow patterns around heart valves: A numerical method, J. Comput.
  Phys. 10(2) (1972) 252--271.

\bibitem{Battista:2015}
N.~A. Battista, A.~J. Baird, L.~A. Miller, A mathematical model and matlab code
  for muscle-fluid-structure simulations, Integr. Comp. Biol.

\bibitem{Peskin:1996}
C.~S. Peskin, D.~M. McQueen, Fluid dynamics of the heart and its valves, in:
  F.~R. Adler, M.~A. Lewis, J.~C. Dalton (Eds.), Case Studies in Mathematical
  Modeling: Ecology, Physiology, and Cell Biology, Prentice-Hall, New Jersey,
  1996, Ch.~14, pp. 309--338.

\bibitem{GriffithThesis:2005}
B.~E. Griffith, Simulating the blood-muscle-vale mechanics of the heart by an
  adaptive and parallel version of the immsersed boundary method (ph.d.
  thesis), Courant Institute of Mathematics, New York University.

\bibitem{Zhang:2014}
C.~Zhang, R.~D. Guy, B.~Mulloney, Q.~Zhang, T.~J. Lewis, The neural mechanism
  of optimal limb coordination in crustacean swimming, PNAS 111 (2014)
  13840--13845.

\bibitem{Hoover:2017}
A.~P. Hoover, B.~E. Griffith, L.~A. Miller, Quantifying performance in the
  medusan mechanospace with an actively swimming three-dimensional jellyfish
  model, J. Fluid. Mech. 813 (2017) 1112--1155.

\bibitem{Miller:2009}
L.~A. Miller, C.~S. Peskin, A computational fluid dynamics of clap and fling in
  the smallest insects, J. Exp. Biol. 208 (2009) 3076--3090.

\bibitem{Kim:2006}
Y.~Kim, C.~S. Peskin, 2d parachute simulation by the immersed boundary method,
  SIAM J. Sci. Comput. 28 (2006) 2294‚Äì2312.

\bibitem{Tytell:2010}
E.~Tytell, C.~Hsu, T.~Williams, A.~Cohen, L.~Fauci, Interactions between
  internal forces, body stiffness, and fluid environment in a neuromechanical
  model of lamprey swimming, Proc. Natl. Acad. Sci. 107 (2010)
  19832‚Äì19837.

\bibitem{Hamlet:2015}
C.~Hamlet, L.~J. Fauci, E.~D. Tytell, The effect of intrinsic muscular
  nonlinearities on the energetics of locomotion in a computational model of an
  anguilliform swimmer, J. Theor. Biol. 385 (2015) 119--129.

\bibitem{Zhu:2011}
L.~Zhu, G.~He, S.~Wang, L.~A. Miller, X.~Zhang, Q.~You, S.~Fang, An immersed
  boundary method by the lattice boltzmann approach in three dimensions,
  Computers and Mathematics with Applications 61 (2011) 3506--3518.

\bibitem{Miller:2012}
L.~A. Miller, A.~Santhanakrishnan, S.~K. Jones, C.~Hamlet, K.~Mertens, L.~Zhu,
  Reconfiguration and the reduction of vortex-induced vibrations in broad
  leaves, J. Exp. Biol. 215 (2012) 2716--2727.

\bibitem{Zhu:2002}
L.~Zhu, C.~S. Peskin, Simulation of a flapping flexible filament in a flowing
  soap film by the immersed boundary method, J. Comp. Phys. 179(2) (2002)
  452--468.

\bibitem{Zhu:2003}
L.~Zhu, C.~S. Peskin, Interaction of two flapping filaments in a flowing soap
  film, Physics of Fluids 15(7) (2003) 1954.

\bibitem{Atzberger:2007}
P.~J. Atzberger, P.~R. Kramer, C.~S. Peskin, A stochastic immersed boundary
  method for fluid-structure dynamics at microscopic length scales, J. Comp.
  Phys. 224(2) (2007) 1255--1292.

\bibitem{Leiderman:2008}
K.~Leiderman, L.~A. Miller, A.~L. Fogelson, The effects of spatial
  inhomogeneities on flow through the endothelial surface layer, J. Theor.
  Biol. 252(2) (2008) 313--325.

\bibitem{Fogelson:2008}
A.~L. Fogelson, R.~D. Guy, Immersed-boundary-type models of intravascular
  platelet aggregation, Comput. Methods Appl. Mech. Engrg. 197 (2008)
  2087‚Äì2104.

\bibitem{Strychalski:2013}
W.~Strychalski, R.~D. Guy, A computational model of bleb formation, Math Med
  Biol. 30(2) (2013) 115--130.

\bibitem{Kramer:2008}
P.~R. Kramer, C.~S. Peskin, P.~J. Atzberger, On the foundations of the
  stochastic immersed boundary method, Comp. Meth. in Appl. Mech. and Eng. 197
  (2008) 2232--2249.

\bibitem{Lee:2010}
P.~Lee, B.~E. Griffith, C.~S. Peskin, The immersed boundary method for
  advection-electrodiffusion with implicit timestepping and local mesh
  refinement, J. Comp. Phys. 229(13) (2010) 5208--5227.

\bibitem{Strychalski:2012}
W.~Strychalski, R.~D. Guy, Viscoelastic immersed boundary methods for zero
  reynolds number flow, Comm. in Comp. Phys. 12 (2012) 462--478.

\bibitem{Du:2014}
J.~Du, R.~D. Guy, A.~L. Fogelson, An immersed boundary method for two-fluid
  mixtures, J. Comp. Phys. 262 (2014) 231--243.

\bibitem{Baird:2015}
A.~J. Baird, L.~D. Waldrop, L.~A. Miller, Neuromechanical pumping: boundary
  flexibility and traveling depolarization waves drive flow within valveless,
  tubular hearts, Japan J. Indust. Appl. Math. 32 (2015) 829--846.

\bibitem{Waldrop:2015}
L.~D. Waldrop, L.~A. Miller, The role of the pericardium in the valveless,
  tubular heart of the tunicate, \emph{{C}iona savignyi}, J. Exp. Biol. 218
  (2015) 2753--2763.

\bibitem{BattistaBioMath:2017}
N.~A. Battista, L.~A. Miller, A fully coupled
  fluid-structure-muscle-electrophysiology model in heart development, BioMath
  Communications Supplement 4(1).

\bibitem{MATLAB:2015a}
MATLAB, version 8.5.0 (R2015a), The MathWorks Inc., Natick, Massachusetts, USA,
  2015.

\bibitem{Python:Python}
G.~{Van Rossum}, Python, version 3.5, https://www.python.org, 2015.

\bibitem{Chorin:1967}
A.~J. Chorin, The numerical solution of the navier-stokes equations for an
  incompressible fluid, Bull. Am. Math. Soc. 73 (1967) 928--931.

\bibitem{Brown:2001}
D.~L. Brown, R.~Cortez, M.~L. Minion, Accurate projection methods for the
  incompressible navier‚Äìstokes equations, J. Comp. Phys. 168 (2001)
  464--499.

\bibitem{Peskin:2002}
C.~S. Peskin, The immersed boundary method, Acta Numerica 11 (2002) 479--517.

\bibitem{Liu:2012}
Y.~Liu, Y.~Mori, Properties of discrete delta functions and local convergence
  of the immersed boundary method, SIAM J. of Numerical Analysis 50 (2012)
  2986--3015.

\bibitem{Cooley:1965}
J.~W. Cooley, J.~W. Tukey, An algorithm for the machine calculation of complex
  fourier series, Math. Comput. 19 (1965) 297--301.

\bibitem{Press:1992}
W.~H. Press, B.~P. Flannery, S.~A. Teukolsky, W.~T. Vetterling, Fast fourier
  transform, Ch. 12 in Numerical Recipes in FORTRAN: The Art of Scientific
  Computing 2 (1992) 490--529.

\bibitem{Lai:2000}
M.~C. Lai, C.~S. Peskin, An immersed boundary method with formal second=order
  accuracy and reduced numerical viscosity, J. Comp. Phys. 160 (2000) 705--719.

\bibitem{Griffith:2009}
B.~E. Griffith, An accurate and efficient method for the incompressible
  navier‚Äì stokes equations using the projection method as a
  preconditioner, J. Comput. Phys. 228 (2009) 7565‚Äì7595.

\bibitem{Bhalla:2013b}
A.~Bhalla, B.~E. Griffith, N.~Patankar, A unified mathematical frame- work and
  an adaptive numerical method for fluid‚Äìstructure interaction with
  rigid, deforming, and elastic bodies, J. Comput. Phys. 250 (2013)
  446‚Äì476.

\bibitem{Stockie:2009}
J.~M. Stockie, Modelling and simulation of porous immersed boundaries,
  Computers and Structures 87 (2009) 701--709.

\bibitem{Whitaker:1986}
S.~Whitaker, Flow in porous media i: A theoretical derivation of darcy's law,
  Transport in Porous Media 1~(1) (1986) 3--25.

\bibitem{Fogelson:1984}
A.~L. Fogelson, A mathematical model and numerical method for studying platelet
  adhesion and aggregation during blood clotting, J. Comput. Phys. 56 (1984)
  111--134.

\bibitem{Fogelson:1992}
A.~L. Fogelson, Continuum models of platelet aggregation: formulation and
  mechanical properties, SIAM JAM 52 (1984) 1089--1110.

\bibitem{Fauci:1993}
L.~Fauci, A.~Fogelson, Truncated newton methods and the modeling of complex
  immersed elastic structures, Commun. Pure Appl. Math 46 (1993) 787‚Äì818.

\bibitem{Fogelson:1999}
N.~T. Wang, A.~L. Fogelson, Computational methods for continuum models of
  platelet aggregation, J. Comp. Phys. 151 (1999) 649--675.

\bibitem{Leiderman:2011}
K.~Leiderman, A.~L. Fogelson, Grow with the flow: a spatial-temporal model of
  platelet deposition and blood coagulation under flow, Math. Med. Biol. 28(1)
  (2011) 47--84.

\bibitem{Leiderman:2014}
K.~Leiderman, A.~L. Fogelson, An overview of mathematical modeling of thrombus
  formation under flow, Thrombosis Research 133 (2014) S12--S14.

\bibitem{Boussinesq:1897}
J.~Boussinesq, Theorie de l`ecoulement tourbillonnant et tumultueux des
  liquides dans les lits rectilignes a grande section (Vol. 1),
  Gauthier-Villars, Paris, France, 1897.

\bibitem{Tritton:1977}
D.~J. Tritton, Physical fluid dynamics, Van Nostrand Reinhold Co., New York,
  NY, USA, 1977.

\bibitem{FFTW05}
M.~Frigo, S.~G. Johnson, The design and implementation of {FFTW3}, Proceedings
  of the IEEE 93~(2) (2005) 216--231, special issue on ``Program Generation,
  Optimization, and Platform Adaptation''.

\bibitem{Ghosh:2015}
S.~Ghosh, J.~Stockie, Numerical simulations of particle sedimentation using the
  immersed boundary method, Commun. Comput. Phys. 18(2) (2015) 380--416.

\bibitem{BGriffithIBAMR}
B.~E. Griffith, \href{https://github.com/IBAMR/IBAMR}{An adaptive and
  distributed-memory parallel implementation of the immersed boundary (ib)
  method} (2014) [cited October 21, 2014].
\newline\urlprefix\url{https://github.com/IBAMR/IBAMR}

\bibitem{Frisani:2012}
A.~Frisani, Direct forcing immersed boundary methods: Finite element versus
  finite volume approach (ph.d. thesis), Nuclear Engineering ,Texas A\&M
  University.

\bibitem{Paraview:2005}
J.~Ahrens, B.~Gerveci, C.~Law, ParaView: An End-User Tool for Large Data
  Visualizations, Elsevier, Atlanta, USA, 2005.

\bibitem{HPV:VisIt}
H.~Childs, E.~Brugger, B.~Whitlock, J.~Meredith, S.~Ahern, D.~Pugmire,
  K.~Biagas, M.~Miller, C.~Harrison, G.~H. Weber, H.~Krishnan, T.~Fogal,
  A.~Sanderson, C.~Garth, E.~W. Bethel, D.~Camp, O.~R\"{u}bel, M.~Durant, J.~M.
  Favre, P.~Navr\'{a}til, {VisIt: An End-User Tool For Visualizing and
  Analyzing Very Large Data}, in: {High Performance Visualization--Enabling
  Extreme-Scale Scientific Insight}, 2012, pp. 357--372.

\bibitem{Chertock:2012}
A.~Chertock, A.~Kurganov, {On Splitting-Based Numerical Methods for
  Convection-Diffusion Equations},
  \url{http://www4.ncsu.edu/~acherto/papers/Chertock-Kurganov.pdf} (2012).

\bibitem{LeVeque:2002}
R.~J. LeVeque, Finite Volume Methods for Hyperbolic Problems, Cambridge
  University Press, Cambridge, UK, 2002.

\bibitem{IBAMR}
B.~E. Griffith, \href{https://github.com/IBAMR/IBAMR}{Ibamr: an adaptive and
  distributed-memory parallel implementation of the immersed boundary method}
  (2014).
\newline\urlprefix\url{https://github.com/IBAMR/IBAMR}

\bibitem{Griffith:IBFE}
B.~E. Griffith, X.~Luo, Hybrid finite difference/finite element version of the
  immersed boundary method, Int. J. Numer. Meth. Engng. 0 (2012) 1--26.

\bibitem{PETSc}
H.~Zhang, \href{http://www.mcs.anl.gov/petsc}{{PETSc}: Portable, extensible
  toolkit for scientific computation} (2009).
\newline\urlprefix\url{http://www.mcs.anl.gov/petsc}

\bibitem{SAMRAI}
{Lawrence Livermore National Laboratory},
  \href{http://www.llnl.gov/CASC/SAMRAI}{{SAMRAI}: Structured adaptive mesh
  refinement application infrastructure} (2007).
\newline\urlprefix\url{http://www.llnl.gov/CASC/SAMRAI}

\bibitem{libMesh}
B.~S. Kirk, J.~W. Peterson, R.~H. Stogner, G.~F. Carey, {\texttt{libMesh}: A
  C++ Library for Parallel Adaptive Mesh Refinement/Coarsening Simulations},
  Engineering with Computers 22~(3-4) (2006) 237--254,
  \url{http://dx.doi.org/10.1007/s00366-006-0049-3}.

\bibitem{openMPI}
E.~Gabriel, G.~E. Fagg, G.~Bosilca, T.~Angskun, J.~J. Dongarra, J.~M. Squyres,
  V.~Sahay, P.~Kambadur, B.~Barrett, A.~Lumsdaine, R.~H. Castain, D.~J. Daniel,
  R.~L. Graham, T.~S. Woodall, Open {MPI}: Goals, concept, and design of a next
  generation {MPI} implementation, in: Proceedings, 11th European PVM/MPI
  Users' Group Meeting, Budapest, Hungary, 2004, pp. 97--104.

\bibitem{Crowl:2011}
L.~M. Crowl, A.~L. Fogelson, Analysis of mechanisms for platelet near-wall
  excess under arterial blood flow conditions, J. Fluid Mech. 676 (2011)
  348--375.

\bibitem{Lewis:2015}
O.~L. Lewis, S.~Zhang, R.~D. Guy, J.~{del Alamo}, Coordination of
  contractility, adhesion and flow in migrating physarum amoebae, J. R. Soc.
  Interface 12(106).

\bibitem{openFOAM:openFOAM}
O.~Foundation, \href{https://github.com/OpenFOAM}{Official openfoam repository}
  (2014).
\newline\urlprefix\url{https://github.com/OpenFOAM}

\bibitem{COMSOL:2011}
C.~Multiphysics, version 4.2 (R2015a), COMSOL Inc., Burlington, MA, USA, 2015.

\bibitem{Fluent:2016}
C.~ANSYS~Fluent, CFD, version 17.0, ANSYS Inc., Cecil Township, Pennslyvania,
  USA, 2016.

\bibitem{Sethian:1999}
J.~A. Sethian, Level Set Methods and Fast Marching Methods : Evolving
  Interfaces in Computational Geometry, Fluid Mechanics, Computer Vision, and
  Materials Science, Cambridge University Press, Cambridge, UK, 1999.

\bibitem{Osher:2002}
S.~J. Osher, R.~Fedkiw, Level Set Methods and Dynamic Implicit Surfaces,
  Springer-Verlag, New York, NY, USA, 2002.

\bibitem{Cortez:2000}
R.~Cortez, M.~Minion, The blob projection method for immersed boundary
  problems, J. Comp. Phys. 161 (2000) 428‚Äì453.

\bibitem{Ubbink:1999}
O.~Ubbink, R.~I. Issa, Method for capturing sharp fluid interfaces on arbitrary
  meshes, J. Comp. Phys. 153 (1999) 26--50.

\bibitem{Udaykumar:2001}
H.~Udaykumar, R.~Mittal, P.~Rampunggoon, A.~Khanna, A sharp interface cartesian
  grid method for simulating flows with complex moving boundaries, J. Comp.
  Phys. 20 (2001) 345--380.

\bibitem{Lee:2003}
L.~Lee, R.~J. Leveque, An immersed interface method for incompressible
  navier-stokes equations, SIAM J. SCI. COMPUT. 25(3) (2003) 832‚Äì856.

\bibitem{Li:2003}
Z.~Li, An overview of the immersed interface method and its applications,
  Taiwanese J. Math. 7(1) (2003) 1--49.

\bibitem{Mittal:2005}
R.~Mittal, C.~Iaccarino, Immersed boundary methods, Annu. Rev. Fluid Mech. 37
  (2005) 239‚Äì261.

\end{thebibliography}

\end{document}